\documentclass[aps,prc,superscriptaddress,showpacs,nofootinbib,floatfix,twocolumn]{revtex4}

\usepackage{doi,graphicx,amssymb,amsmath,dcolumn,hyperref}
\newcommand{\unit}[1]{\ensuremath{\, \mathrm{#1}}}
\newcommand{\etal}{\textit{et al.}}
\newcommand{\ie}{\textit{i.e.}}

\newcolumntype{d}[1]{D{.}{.}{#1} }
\usepackage[percent]{overpic}
\begin{document}

\title{Nuclear Waffles}
\author{A. S. Schneider}\email{andschn@indiana.edu}
\affiliation{Department of Physics and Nuclear Theory Center, Indiana University, Bloomington, IN 47405, USA}
\author{D. K. Berry}\email{dkberry@iu.edu}
\affiliation{University Information Technology Services, Indiana University, Bloomington, IN 47408, USA}
\author{C. M. Briggs}\email{briggchm@indiana.edu}
\author{M. E. Caplan}\email{mecaplan@indiana.edu}
\author{C. J. Horowitz}\email{horowit@indiana.edu}
\affiliation{Department of Physics and Nuclear Theory Center, Indiana University, Bloomington, IN 47405, USA}

\date{\today}

\begin{abstract}

\begin{description}
 \item[Background] The dense neutron-rich matter found in supernovae and inside neutron stars is expected to form complex nonuniform phases, often referred to as \textit{nuclear pasta}. The pasta shapes depend on density, temperature and proton fraction and determine many transport properties in supernovae and neutron star crusts.
 \item[Purpose] To characterize the topology and compute two observables, the radial distribution function (RDF) $g(r)$ and the structure factor $S(q)$, for systems with proton fractions $Y_p=0.10,\,0.20,\,0.30$ and $0.40$ at about one third of nuclear saturation density, $n=0.050\unit{fm}^{-3}$, and temperatures near $kT=1\unit{MeV}$.
 \item[Methods] We use two recently developed hybrid CPU/GPU codes to perform large scale molecular dynamics (MD) simulations with 51\,200 and 409\,600 nucleons. From the output of the MD simulations we obtain the two desired observables.
 \item[Results] We compute and discuss the differences in topology and observables for each simulation. We observe that the two lowest proton fraction systems simulated, $Y_p=0.10$ and $0.20$, equilibrate quickly and form liquid-like structures. Meanwhile, the two higher proton fraction systems, $Y_p=0.30$ and $0.40$, take a longer time to equilibrate and organize themselves in solid-like periodic structures. Furthermore, the $Y_p=0.40$ system is made up of slabs, \textit{lasagna phase}, interconnected by defects while the $Y_p=0.30$ systems consist of a stack of perforated plates, the \textit{nuclear waffle} phase. 
 \item[Conclusions] The periodic configurations observed in our MD simulations for proton fractions $Y_p\ge0.30$ have important consequences for the structure factors $S(q)$ of protons and neutrons, which relate to many transport properties of supernovae and neutron star crust. A detailed study of the waffle phase and how its structure depends on temperature, size of the simulation and the screening length showed that finite-size effects appear to be under control and, also, that the plates in the waffle phase merge at temperatures slightly above 1.0\unit{MeV} and the holes in the plates form an hexagonal lattice at temperatures slightly lower than 1.0\unit{MeV}.
\end{description}
% In this work we continue our study of complex nonuniform phases of nuclear matter using large scale molecular dynamics simulations.
% Our simulations are performed by two recently developed hybrid CPU/GPU codes and use as many as 409\,600 nucleons.
% We characterize the topology and compute the radial distribution function (RDF) $g(r)$ and the structure factor $S(q)$ for 
% systems with proton fractions of $Y_p=0.10,\,0.20,\,0.30$ and $0.40$ at about one third of nuclear saturation density ($n=0.050\unit{fm}^{-3}$).
% We observe that the structure factors of the two lowest proton fraction systems, $Y_p=0.10$ and $0.20$, have a liquid-like structure
% while the higher proton fractions, $Y_p=0.30$ and $0.40$, exhibit diffraction peaks characteristic of solid-like structures.
% Our focus is the $Y_p=0.30$ systems as they have the least usual structure: a stack of perforated plates.
% We study the stability of these perforated plates, and how their structure depends on temperature, size of the simulation and the screening length used. 
% The plates merge at temperatures slightly above 1.0\unit{MeV} and the holes in the plates form an hexagonal lattice at lower temperatures.
\end{abstract}

% PACS numbers
% 26.60.-c Nuclear matter aspects of neutron stars
% 26.60.Dd Neutron stars cores
% 26.50.+x Nuclear physics aspects of Supernovae
% 64.70.M- Transitions in liquid crystals
% 02.70.Ns Molecular dynamics and particle methods

\maketitle

\section{Introduction}\label{sec:Intro}

It is widely accepted that dense neutron-rich matter forms during a core-collapse supernova and exists between the crust and the core of a neutron star.
A combination of theoretical arguments and numerical simulations suggests that this type of matter forms complex nonuniform structures, nowadays referred to as \textit{nuclear pasta}.
These complex nonuniform structures form because the system is unable to minimize all its fundamental interactions \cite{Pethick19987}.
Here the interactions are the attractive short-range nuclear force, $\mathcal{O}\sim 1\unit{fm}$, and the repulsive long-range Coulomb force, $\mathcal{O}\sim 10-100\unit{fm}$.
There is an ongoing effort aiming to determine the possible shapes of the pasta, its phase-transitions and their properties 
as these are relevant to the equation of state of nuclear matter \cite{PhysRevLett.41.1623}, 
neutrino opacities in supernovae \cite{PhysRevC.69.045804,PhysRevC.83.035803} and electric transport in the neutron star crust \cite{pons2013highly}.

Often, studies of nuclear pasta make use of symmetry arguments to determine what is the most favored structure at a given density, temperature and proton fraction.
For example, some mean-field calculations solve the equations of motion of dense matter in a Wigner-Seitz approximation in one, two and three dimensions
and choose the favored geometry as the one that minimizes the energy density of the system for a given density and proton fraction 
\cite{PhysRevC.72.015802,PhysRevC.78.015802,PhysRevC.79.035804,PhysRevC.87.028801,PhysRevC.89.045807}.
Other works based on liquid-drop models and Thomas-Fermi approximation also have explicit assumptions about the geometrical shapes of nuclear pasta 
\cite{PhysRevLett.50.2066,PTP.71.320,1984PThPh..72..373O,Watanabe2000455}.
As noted by Williams and Koonin in Reference~\cite{Williams1985844} these symmetry arguments limit the possible structures to uniform 
and five geometries: spheres (3D), cylinders (2D), plates (1D), tubes (2D) and bubbles (3D).
Thus, they performed simulations without any \textit{a priori} assumption on the pasta geometry and 
were able to show that the five assumed geometries are good descriptions of nuclear matter in certain density ranges.

The way the primitive cells are stacked in small volume simulations limits the configurations of the 3D phases to simple cubic lattices while the 2D phases can only form a square lattice.
With this in mind Oyamatsu \etal\, added extra configurations in which the primitive cells could be stacked to include bcc and fcc lattice configurations to the 3D phases
as well as an hexagonal lattice configuration to the two-dimensional phases \cite{1984PThPh..72..373O}.
Though their calculations still had assumptions on the nuclear structures formed within their unit cells 
they concluded that for symmetric nuclear matter the 3D phases prefer to form bcc lattices while the 2D phases form hexagonal lattices.
Recently Okamoto \etal\, used Thomas-Fermi approximation to determine pasta structures within a volume large enough 
to include more than a single periodic image of its structure \cite{Okamoto2012284}.
Their computations, which had no assumptions about the pasta geometries or lattice structures formed, 
indicated that symmetric nuclear matter can transition from bcc to fcc structures in the three-dimensional phases and from a honeycomb to a square lattice in the two-dimensional phases.
For lower proton fractions, $Y_p=0.10$ and $0.30$, they demonstrated that the fcc and simple cubic structures are favoured.
Furthermore, in their search for the ground-state they found more exotic pasta shapes, albeit in a metastable state. 
Amongst the geometries obtained were dumbbell-like and diamond-like structures, as well as coexistence of phases of different dimensionalities, 
for instance mixtures of droplets and rods appeared at low densities and slabs and tubes at intermediate densities.
Other works explored structures such as gyroid and double-diamond morphologies \cite{PhysRevLett.103.132501,PhysRevC.83.065811} 
as the existence of these exotic shapes may have important implications to the pasta properties.

Advances in computational power in the past decade have allowed for sophisticated calculations beyond mean-field, Thomas-Fermi and liquid drop model approximations. These include fully self-consistent calculations using a Skyrme-Hartree-Fock+BCS calculation at finite temperature \cite{PhysRevC.79.055801, PhysRevLett.109.151101} and time-dependent Hartree-Fock simulations \cite{PhysRevC.87.055805, Schuetrumpf:2014aea,Schutrumpf:2014vqa}. These computations showed a richer variety of pasta shapes than the five geometries typically reproduced. However, due to their complexity, these calculations are often limited to a single periodic structure so that the pasta shapes obtained may exhibit significant dependence on the finite-size of the simulation. In fact, it was recently showed by Molinelli \etal\, for molecular dynamics simulations of about less than 10\,000 nucleons that the pasta shapes may differ based on the geometry chosen for the simulation volume \cite{Molinelli:2014uta}. Therefore, it is necessary to perform simulations with a much larger number of nucleons so that finite-size effects can be overcome.

Because of limitations in computational power calculations with more than a few thousand nucleons are only manageable by considerably simplifying nucleon interactions. That can be attained by replacing the nucleon interactions by schematic forces that reproduce some of the properties of finite nuclei and nuclear matter, even if that implies ignoring shell effects and other important physics. This is what is done in works that study nuclear pasta using semi-classical molecular dynamics (MD) \cite{PhysRevC.69.045804, PhysRevC.70.065806, PhysRevC.72.035801, PhysRevC.78.035806, PhysRevC.86.055805, GiménezMolinelli201431, PhysRevC.88.065807, PhysRevC.89.055801}, quantum molecular dynamics (QMD) \cite{PhysRevC.57.655,PhysRevC.66.012801,PhysRevC.68.035806,PhysRevC.69.055805,PhysRevC.77.035806,PhysRevLett.94.031101,PhysRevLett.103.121101} and Monte-Carlo methods \cite{PhysRevC.88.025807,PhysRevC.85.015807}. So far, the largest simulations reported in literature were performed by Horowitz \etal\, and included up to 100\,000 nucleons though it is not clear if those simulations were run for long enough for the system to equilibrate \cite{PhysRevC.70.065806,PhysRevC.72.035801}.

In a previous paper we studied nuclear pasta formation using MD \cite{PhysRevC.88.065807}. In that work we evolved dense matter with a proton fraction of $Y_p=0.40$ at a temperature of 1\unit{MeV} 
from high to low densities, $n=0.10\unit{fm}^{-3}$ to $n\sim0.01\unit{fm}^{-3}$, by expanding the simulation volume at different rates. We explicitly observed the nucleation mechanism as the pasta transitioned from one phase to the next and quantified the topologycal transitions by calculating Minkowski functionals on a suitable isosurface of the structures formed. Specifically, we noted that once the density reached approximately $n=0.04\unit{fm}^{-3}$ the system transitioned from plates (``lasagna'' phase) to cylinders (``spaghetti phase''). During the transition, holes appeared in the lasagna plates and a phase similar to perforated plates or cross linked network of spaghetti formed. As density was decreased further the cross links disappeared to produce isolated nearly straight spaghetti strands. Study of finite size effects of this phase of perforated plates is the main focus of this work. This is done running MD simulations of 51\,200 and 409\,600 nucleons at a density of $n=0.05\unit{fm}^{-3}$ and proton fraction of $Y_p=0.30$ for up to $15\times10^6$ MD time steps. This proton fraction is slightly lower than in our previous work and was chosen as the cross-linked phase was more stable. Besides that it also allows us to compare the topology of our results to the work of Pais and Stone \cite{PhysRevLett.109.151101} and Schuetrumpf \etal\, \cite{PhysRevC.87.055805,Schutrumpf:2014vqa} as they obtained a similar phase at similar proton fractions and densities. We also note here that Sebille \etal\, using a dynamic self-consistent mean-field model also obtained a similar phase of stacked perforated plates at this same density for symmetric nuclear matter \cite{PhysRevC.84.055801}.

One of the purposes of performing large nuclear pasta simulations is to determine 
how the pasta phases affect observable quantities present in supernovae and neutron stars.
While nucleon clustering and long range order of the pasta structures are relevant for neutrino pasta-scattering \cite{PhysRevC.69.045804}, 
impurities and/or defects may be important for heat and electrical conductivity and pulsar spin-down \cite{pons2013highly,PhysRevLett.93.221101}.
Because MD allows for much larger simulations than possible with quantum calculations and it is straightforward to
track the time evolution of the system, we can directly calculate observables like radial distribution function (RDF), $g(r)$, and 
its Fourier transform, the static-structure factor $S(q)$.
Therefore, besides the study of the perforated plates phase, we also calculate the topology and observables of structures 
formed at a density of $n=0.050\unit{fm}^{-3}$ for four proton fractions, $Y_p=0.10,\,0.20\,0.30$ and $0.40$ and compare their properties.
Time and frequency dependent observables may also be computed and will be the topic of a future work. 

This manuscript is arranged as follows. In Section~\ref{ssec:Formalism} we review our MD formalism while Section~\ref{ssec:codes} is devoted to the CPU/GPU codes used in our simulations. Afterwards, in Section~\ref{sec:Results} we present our results. The section starts with a discussion of four 51\,200 nucleon simulations with different proton fractions, Section \ref{ssec:Yp}. We then move our focus to simulations with proton fraction $Y_p=0.30$ of different sizes and screening lengths, Section~\ref{ssec:30}, and finish with a discussion of some observables that can be obtained from the MD simulations, Section~\ref{ssec:Obs}.
Finally, we conclude in Sec. \ref{sec:Conclusions}.

\section{MD Code}\label{sec:Formalism}

We start this section discussing the formalism used in our codes, Sec. \ref{ssec:Formalism}, a brief description of how we obtain the relevant Minkowski functionals, Sec. \ref{ssec:minkowski}
and then describe how the CPU/GPU codes work in Sec. \ref{ssec:codes}.

\subsection{Formalism}\label{ssec:Formalism}

Following is a review of our MD formalism, as it is the same as the one used by Horowitz \etal\, and others in previous works 
\cite{PhysRevC.69.045804,PhysRevC.70.065806,PhysRevC.72.035801,PhysRevC.78.035806,PhysRevC.86.055805,GiménezMolinelli201431,PhysRevC.88.065807,PhysRevC.89.055801}.
We use a cubic box with periodic boundary conditions to simulate systems of neutrons and protons immersed in a degenerate relativistic free Fermi electron gas.
The nucleons are mass $M=939\unit{MeV}$ point-like particles that interact via two-body potentials of the form
\begin{subequations}
\begin{align}
 V_{np}(r)&=a e^{-r^2/\Lambda}+[b-c]e^{-r^2/2\Lambda}\\
 V_{nn}(r)&=a e^{-r^2/\Lambda}+[b+c]e^{-r^2/2\Lambda}\\
 V_{pp}(r)&=a e^{-r^2/\Lambda}+[b+c]e^{-r^2/2\Lambda}+\frac{\alpha}{r}e^{-r/\lambda}.
\end{align}
\end{subequations}
The $n$ and $p$ indexes denote whether the potential is for a neutron-proton, neutron-neutron or proton-proton interaction. In the equations above, $r$ is the distance between the two nucleons and $a$, $b$, $c$ and $\Lambda$ are constants adjusted to approximately reproduce some bulk properties of pure neutron matter and symmetric nuclear matter as well as the binding energies of selected nuclei \cite{PhysRevC.69.045804}. Their values are given in Table \ref{Tab:parameters}. As there have been studies on the dependence of the pasta phases on the density dependence of the nuclear symmetry energy, for an example see Reference~\cite{PhysRevC.89.045807}, we quote our value for this quantity: $S=40.7\unit{MeV}$. We also obtain a value of $K=372\unit{MeV}$ for the nuclear compressibility, although we do not expect our results to be very sensitive to this somewhat high value.

The proton-proton interaction $V_{pp}$ also has a term proportional to the fine-structure constant $\alpha$. This is the Coulomb repulsion between protons screened by the background electron gas. The screening has a characteristic length $\lambda$ that depends on the electron Fermi momentum $k_F=(3\pi^2n_e)^{1/3}$, where $n_e$ is the electron density and the electron mass $m_e$.
Its value is
\begin{equation}
 \lambda=\frac{\pi^{1/2}}{2\alpha^{1/2}}\left(k_F\sqrt{k_F^2+m_e^2}\right)^{-1/2}
\label{eq:lambda}
\end{equation}
In most previous works $\lambda$ was fixed to an arbitrary value $\lambda=10\unit{fm}$.
Though we do that in Section~\ref{ssec:Yp}, in Section~\ref{ssec:30} we compare our results for runs with both $\lambda=10\unit{fm}$ and $\lambda=\lambda_{TF}$ given by Eq. \eqref{eq:lambda}, \ie\, $\lambda=13.6\unit{fm}$.

\begin{table}[h]
\caption{\label{Tab:parameters} Nuclear interaction parameters. The parameter $a$ defines the strength of the short-range repulsion between nucleons, $b$ and $c$ the strength of their intermediate-range attraction and $\Lambda$ the length scale of the nuclear potential.}
\begin{ruledtabular}
\begin{tabular}{*{4}{c}}
$a$ (MeV) &$b$ (MeV)&$c$ (MeV)&$\Lambda$ (fm$^{2}$) \\
\hline
  110     &  $-$26    &   24    &    1.25      \\
\end{tabular}
\end{ruledtabular}
\end{table}

\subsection{Minkowski functionals}\label{ssec:minkowski}

To quantify the shapes of the structures formed in our simulations we use Minkowski functionals. 
In three dimensions any shape may be classified in terms of four Minkowski functionals: volume $V$, area $A$, mean breath $B$ and Euler characteristic $\chi$.
In our simulations the occupied volume $V$ is defined by the region enclosed by a nuclear surface of total area $A$.
Meanwhile, the mean breadth $B$ and Euler characteristic $\chi$ are, respectively, proportional to the surface integrals of 
the mean curvature $\tfrac{1}{2}(\kappa_\text{min}+\kappa_\text{max})$ and the Gaussian curvature $(\kappa_\text{min}\kappa_\text{max})$.
Here $\kappa_\text{min}$ and $\kappa_\text{max}$ are the minimum and maximum values for the curvature on each point of the surface.
Furthermore, the Gaussian curvature may be related to the number of structures or the connectivity of the shapes formed \cite{Michielsen2001461}.

As in our previous work, Reference~\cite{PhysRevC.88.065807}, the nuclear surface is defined as isosurfaces of charge density $n_{\rm{ch}}=0.03\unit{fm}^{-3}$ 
obtained by folding a three-dimensional unitary Gaussian with standard deviation of $\sigma=1.5\unit{fm}$ around each proton of the system. 
The surface integrals were performed using the prespcription of Lang \etal\,\cite{lang01}.
To track the evolution of a system and to directly compare the topology of simulations of different sizes
we calculate the average mean curvature, $B/A$, and the average Gaussian curvature, $\chi/A$.

\subsection{GPU codes}\label{ssec:codes}

The most time consuming task when solving the equations of motion of the system described above is the computation of the forces acting on each nucleon. In this work we use an upgraded version of the Indiana University Molecular Dynamics (IUMD) Fortran code used in our previous paper, Reference~\cite{PhysRevC.88.065807}. Amongst the upgrades are a neighbor-list scheme to calculate the nuclear forces using CPUs and the use of GPUs, whenever available, to calculate the long-range Coulomb interaction between protons. The details of the code are described in Sec. \ref{sssec:IUMD}. We also describe another newly developed Fortran code, \textsc{CubeMD}, which also makes use of CPUs and GPUs. This code is discussed in Sec. \ref{sssec:CubeMD}. In a forthcoming paper we will discuss the performance of each code as it depends on density, temperature and proton fraction of the simulation.

% Both of these codes are a significant improvement over the previous versions of the IUMD code.
% One of its main advantage is that the nuclear and Coulomb forces computations are split between CPUs and GPUs.
% Since the nuclear force falls off to zero rather quickly with distance, up to machine precision, 
% this component of the force on any nucleon depends at most on the positions of other nucleons within a range of about 11\unit{fm} of it.
% Thus, one can efficiently use CPUs to calculate which particles are within a close range of each nucleon.  
% This can be done by keeping track of a neighbor list, as it is done on the IUMD code, 
% or by dividing the simulation volume into small cubes, as it is done in CubeMD, and only computing pairwise distances of nucleons within the same box or adjacent boxes.
% While the CPUs compute the nuclear forces the long range Coulomb force, which only needs to be computed amongst proton pairs, is obtained by the GPUs.
% In Sec. \ref{sssec:IUMD} we describe some of the details of the latest version of the IUMD code while in Sec. \label{sssec:CubeMD} we describe details of the CubeMD code.

\subsubsection{The IUMD code}\label{sssec:IUMD}

The IUMD code has been developed for the past decade and has recently undergone a major reformulation
to take full advantage of the \textsc{Big Red II} supercomputer acquired by Indiana University last
year. \textsc{Big Red II} is a Cray XE6/XK7. The XE6 part of the machine consists of 344 dual CPU
compute nodes, where each CPU is an Advanced Micro Devices 16-core Abu Dhabi Opteron. Each of these
nodes has 64 GB of RAM. The XK7 part consists of 676 CPU/GPU compute nodes, each containing one
16-core AMD Interlagos Opteron CPU, one Nvidia Kepler K20 GPU, and 32 GB of RAM \cite{BigRed2}. IUMD
is a parallel code that can run on either the dual CPU nodes, or the CPU/GPU nodes, using MPI
(Message Passing Interface) to communicate between nodes, OpenMP threads on the 16 cores of each CPU,
and Portland Group CUDA Fortran on each GPU.  IUMD takes full advantage of the compute power of
CPU/GPU nodes by calculating nuclear forces on the CPUs while computing the Coulomb forces on the
GPUs via a straightforward particle-particle algorithm.  On CPU-only nodes of \textsc{Big Red II},
and any other machines that do not have hybrid CPU/GPU architecture it is also possible to run the
code using only CPUs.

Decomposition of the force calculation among compute nodes is best understood by thinking of all the
two-particle interactions as making up a \emph{force matrix}. Element $ij$ of the matrix corresponds
to the force $\boldsymbol{f}_{ij}$ that \emph{source} particle $j$ exerts on \emph{target} particle $i$.
Of course, sources and targets are the same $N$ particles overall, but thinking of them as sources
acting on targets simplifies explanation. In the parallel code, the force matrix is decomposed into
$P$ block rows and $Q$ block colums, where $PQ$ is the total number of MPI processes (one process per
compute node). Each process is assigned one block, and is responsible for calculating the action of
its $N/Q$ sources on its $N/P$ targets.  In order to simplify communication between processes, as
well as the Coulomb calculation on the GPUs, IUMD does not use Newton's third law to calculate the
reaction of targets on sources. This decomposition resembles a customary cell algorithm, except that
the cells are abstract rather than a geometrical division of real space. Once assigned to a process,
particles stay there; they do not need to be moved from process to process as they would in a
spatially based cell algorithm.

After all processes have calculated the forces their sources exert on their targets, forces are
summed along the $Q$ processes of each row to get the total force on each target.  This is done
by an MPI \emph{allreduce} which leaves each process with the total force on each of its targets.
Note that these are row-wise allreduces, so that in principle the $P$ allreduces can be done
concurrently. Thus the code should scale to very large node counts on machines that can actually
do them concurrently.  A time step is finished by each process applying a velocity Verlet update
to its targets, followed by another set of allreduces, this time along each block column, to copy
the new target positions to sources belonging to that column. Since each column has the complete
set of new target positions, these allreduces can also be done concurrently. The only time an
allreduce over all processes is required, is when calculating total potential energy, or virial
for the pressure. However, these calculations are needed relatively infrequently.

Each MPI process calculates Coulomb forces by a simple particle-particle algorithm. All source
and target proton positions are sent to the GPU, which sums the force of all sources on each
target, and returns the forces to the CPU. The GPU version of the code does not use a cut-off
or other work reducing measure, so the Coulomb calculation has computational complexity
$\mathcal{O}((Y_p N)^2/(PQ))$.  In the CPU only version the performance of the Coulomb force
calculations can be improved by setting a cut-off to the Coulomb interaction, though this cut-off
still has to be large enough to allow distant protons to interact with each other. More details
of the Coulomb force calculations using GPUs or CPUs were described in Reference~\cite{berry2013experiences}.
While the GPU calculates the Coulomb interaction, the CPU calculates nuclear forces via a cell and
neighbor list algorithm. Since the nuclear force has a range of only few fermi, at most a few
thousand source nucleons will be within range of each target nucleon, even at saturation density.
Because sources are randomly distributed among $Q$ nodes of each block row of the force matrix,
this is reduced to perhaps hundreds per target per node, making a neighbor list algorithm very
efficient.

In detail, the code builds a neighbor list $L_i$ of all sources within a distance $r_{nuc}+\delta r_{nuc}$
of target $i$. The force on $i$ is calculated only from its interaction with sources in $L_i$ that
are within distance $r_{nuc}$. We set $r_{nuc}=11.5\unit{fm}$ in all runs reported in this paper,
as the nuclear force drops well below machine precision by this distance, even for IEEE 64-bit
arithmetic.  We could probably reduce $r_{nuc}$ to 9 or 8 fm, but took a conservative approach for
these runs.  Sources are included in $L_i$ from the buffer zone of thickness $\delta r_{nuc}$ about
the interaction sphere so lists do not have to be rebuilt as nucleons move in and out of interaction
range. Rather, list $L_i$ needs to be rebuilt only when the distance $i$ has moved, plus the
maximum distance any source on a node has moved since the last build is greater than $\delta r_{nuc}$.
We have found it more efficient to rebuild all lists on all nodes when any one of them needs rebuilding,
as the list-building procedure takes some time, and interrupts flow of the simulation. This requires
an MPI allreduce of a logical variable from each process telling whether it needs to do a rebuild.
For the size of runs we have done, this all-process allreduce is not too time-consuming, and results
in more efficient runs. However in principle, the decision to rebuild lists only needs to be done on a
process-by-process basis.

As just described neighbor list builds would be of $\mathcal{O}(N^2/PQ)$ computational complexity,
as distances between all targets and all sources on a node must be checked. This complexity is
reduced considerably by coupling the algorithm with a cell algorithm.  Each process divides the
whole simulation volume into cells of width $r_{nuc}+ \delta r_{nuc}$, and figures out which cell
each source is in.  This is an order $\mathcal{O}(N/Q)$ operation. Then for each target $i$, only
$i$'s cell and its 26 neighboring cells must be checked in order to build $L_i$.  Note that this
requires no communication between processes.  Even though this reduces work required to build the
$L_i$, builds should still be done as infrequently as possible, implying $\delta r_{nuc}$ should
be large.  However the number of sources in each list grows as $(r_{nuc}+ \delta r_{nuc})^3$, so
$r_{nuc}+ \delta r_{nuc}$ should be kept small.  We have chosen $\delta r_{nuc}=4.0\unit{fm}$, as
a good trade-off of list size vs. frequency of builds. For $Q=1$ this would result in about $3\,200$
sources in each list for density $n=0.20 \unit{fm}^{-3}$, well above saturation density. Of these,
only about $1\,350$ would be within interaction range $r_{nuc}$. Note that for parallel runs these
numbers are reduced by the number $Q$ of MPI processes in each row.  For the densities and
temperatures we usually consider in our works the neighbor lists are rebuilt every dozen or so time
steps depending on how close to equilibrium the system is.

The algorithm just described is a significant improvement over the one used in our previous paper
where the distance over every pair of particles had to be calculated and the code would scale with
$\mathcal{O}(N^2)$.

\subsubsection{The \textsc{CubeMD} code}\label{sssec:CubeMD}

The \textsc{CubeMD} code is also a hybrid CPU/GPU code that works similarly to the IUMD code.
It calculates the nuclear forces on the CPUs while the GPUs take care of the Coulomb interactions amongst the protons. 
The difference is in how the nuclear forces are calculated; while the IUMD code builds neighbor lists for each nucleon
the \textsc{CubeMD} code divides the simulations volume into cubes of sides of approximately 4\unit{fm}.
Each nucleon is then tagged with a number that specifies which of the smaller cubes it belongs to.
The force on a target nucleon is computed only for the potential due to source nucleons in the same cube as the target or in adjacent ones.
The adjacent cubes are determined in such way as to preserve periodic boundary conditions.
We note that as of now the \textsc{CubeMD} code uses only a single CPU/GPU compute node. 
Its performance is slightly better than the IUMD code running on a single compute node.

% \subsubsection{Performance}\label{sssec:Perf}
% [Add a few tables about the codes performance.]

\section{Results}\label{sec:Results}

In this section we describe our simulations and what we have learned from them.
Though all the results presented here are from simulations performed with the \textsc{IUMD} code we did obtain very similar results with the \textsc{CubeMD} code.
However, we decided to omit those results from this work to make the presentation of our results clearer.

We start in Sec. \ref{ssec:Yp} with a comparison of the topologies of systems evolved at a constant density of 
$n=0.050\unit{fm}^{-3}$ and temperature $kT=1.0\unit{MeV}$ for different proton fractions.
The topology is characterized by the average mean $B/A$ and Gaussian $\chi/A$ curvatures \cite{Michielsen2001461}.
In Sec. \ref{ssec:30} we focus on systems with proton fraction $Y_p=0.30$.
We compare how the average mean and Gaussian curvatures evolve for simulations of 51\,200 and 409\,600 nucleons 
from different initial conditions, sizes and screening length and discuss their topological structures.
Finally we finish Sec. \ref{ssec:Obs} discussing how we obtain the radial distribution functions (RDFs) $g(r)$ and the structure factors $S(q)$ from MD simulations.

\subsection{Systems of different proton fractions}\label{ssec:Yp}

We start this section discussing the topologies of systems of different proton fractions.
Using the IUMD code we simulated systems with 51\,200 nucleons in a cubic box with nucleon number density $n=0.050\unit{fm}^{-3}$, 
temperature $kT=1.0\unit{MeV}$ and proton fractions $Y_p=0.10,\,0.20,\,0.30$ and $0.40$.
For the simulations discussed in this section we fixed the screening length to $10\unit{fm}$.
Since a constant density of $n=0.050\unit{fm}^{-3}$ implies a box with length size $100.8\unit{fm}$
the ratio of box length $L$ to screening length $\lambda$ is approximately 10.
Had we used the screening length $\lambda_{TF}$ obtained in the relativistic Thomas-Fermi approximation
the ratio of box length $L$ to screening length $\lambda_{TF}$ would be somewhat smaller, see Table \ref{tab:comp}, and increase with lower proton fractions.

\begin{table}[h]
\caption{\label{tab:comp} Comparison of screening length $\lambda=10\unit{fm}$ used in the simulations and 
the relativistic Thomas-Fermi screening $\lambda_{TF}$ to the box length $L=100.8\unit{fm}$.} 
\begin{ruledtabular}
\begin{tabular}{D{.}{.}{1.2} D{.}{.}{2.2} D{.}{.}{1.2} D{.}{.}{1.2}}
\multicolumn{1}{c}{$Y_p$} & \multicolumn{1}{c}{$\lambda_{TF}\unit{(fm)}$} & \multicolumn{1}{c}{$L/\lambda_{TF}$} & \multicolumn{1}{c}{$L/\lambda$}  \\
\hline
0.10 &  19.610 & 5.14 & 10.08 \\
0.20 &  15.565 & 6.48 & 10.08 \\
0.30 &  13.597 & 7.41 & 10.08 \\
0.40 &  12.354 & 8.16 & 10.08 \\
\end{tabular}
\end{ruledtabular}
\end{table}

From Equation~\eqref{eq:lambda} and the results in Table \ref{tab:comp} we note that, for electrically neutral ultra-relativistic systems ($k_F\gg m_e$) such as the ones where nuclear pasta forms, the Thomas-Fermi screening lengths is proportional to $Y_p^{-1/3}$, \ie\, $\lambda_{TF}\propto Y_p^{-1/3}$.
We also note that in the worst case scenario presented above, $Y_p=0.10$, the value of $\lambda=10\unit{fm}$ 
is within a factor of two of the screening predicted by the Thomas-Fermi approximation.
These values for the screening $\lambda_{TF}$ are much smaller than the ones estimated by Alcain \etal\, in 
Reference~\cite{PhysRevC.89.045807} using a non-relativistic approximation, $m_e>>k_F$ in Eq. \ref{eq:lambda}.
In their work they simulated isospin symmetric nuclear matter which, 
in a relativistic approximation ($m_e<<k_F$), implies a screening length $\lambda_{TF}\sim11.5\unit{fm}$ at the density used in this work, $n=0.050\unit{fm}^3$.
Thus, following their conclusions we expect that for large simulations such as the ones presented here, 
a screening of $10\unit{fm}$ should be sufficient to at least correctly predict the signs for 
the average mean and Gaussian curvatures of the systems with higher proton fractions, $Y_p\gtrsim0.30$.
The differences between the predictions of fixing $\lambda$ for $Y_p=0.30$ and using the Thomas-Fermi approximation will be explored in Section \ref{ssec:30}.

\begin{figure}[ht]
\centering
\includegraphics[width=0.5\textwidth]{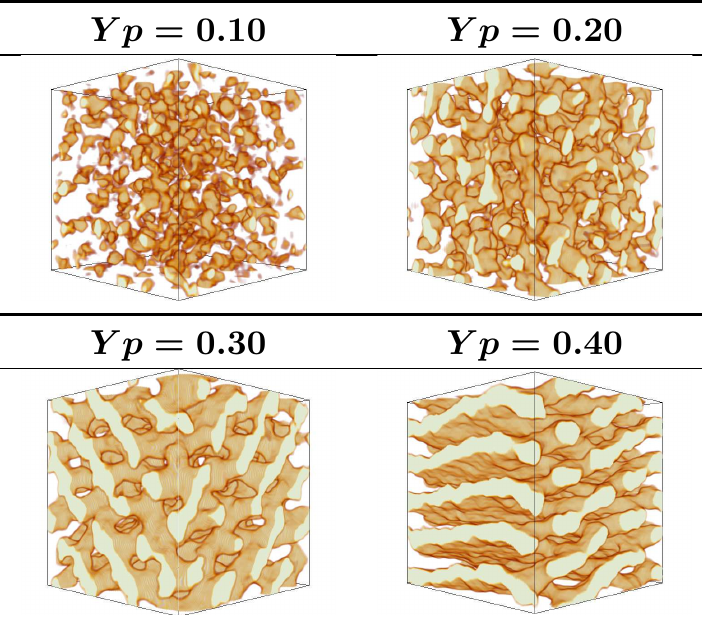}
\caption{\label{fig:Yp} (Color on line) Charge density isosurfaces of runs with 51\,200
    nucleons, mean density $n=0.05\unit{fm}^{-3}$, temperature
    $kT=1.00$ MeV, and proton fractions $Y_p = 0.10, 0.20, 0.30$ and
    0.40 after $10^7$ fm/c evolution time. In this figure, and all
    similar ones throughout this paper, the golden surfaces represent
    isosurfaces of charge density $n_{\rm{ch}}=0.03\unit{fm}^{-3}$,
    while the cream color shows regions such that $n_{\rm{ch}}>0.03
    \unit{fm}^{-3}$. All such figures were generated using ParaView \cite{Paraview}.}
\end{figure}

Each simulation described in this section was evolved for $10^7\unit{fm/c}$ in time steps of $2\unit{fm/c}$.
The final configuration of each simulation is shown in Figure~\ref{fig:Yp}.
To generate Figure 1, a gaussian of unit volume was folded about each proton. These gaussians were then summed at each point of a fine 3D grid overlaying the simulation volume, and an isosurface corresponding to charge density $n_{\rm{ch}}=0.03\unit{fm}^{-3}$ constructed. Details of this construction are given in Reference~\cite{PhysRevC.88.065807}.
We see that the lowest proton fraction, $Y_p=0.10$, formed a phase that consist of small deformed nuclei
while the $Y_p=0.20$ system is mostly formed of deformed elongated nuclei that resemble the spaghetti phase.
The two larger proton fractions, $Y_p=0.30$ and $0.40$, formed structures that spread along the whole length of the simulation volume; 
the $Y_p=0.40$ proton fraction formed flat sheets interconnected by defects, while the $Y_p=0.30$ proton fraction formed perforated plates we named \textit{nuclear waffles}.
The waffle phase is the subject of the following section while the defects in pasta structure will be the subject of a forthcoming paper.

We quantify the shapes formed by calculating the Minkowski functionals (area, mean curvature and Gaussian curvature) of the charge isosurface of density $n_{\rm{ch}}=0.030\unit{fm}^{-3}$. 
We calculate these quantities the same way as in Reference~\cite{PhysRevC.88.065807}, and refer the reader to that paper for details. 
The reason we evolved our simulations for $10^7$ fm/c was that that was the time the slowest converging run took to appear to equilibrate. 
While the Minkowski functionals of the $Y_p=0.10$ and $0.20$ runs stopped evolving after about $2\times10^5\unit{fm/c}$, the $Y_p=0.40$ run took about $2\times10^6\unit{fm/c}$ to reach equilibrium. The slowest converging run was the $Y_p=0.30$. The Minkowski functionals took about $5\times10^6\unit{fm/c}$ to reach an apparent asymptotic value. 
In Table \ref{tab:mf1} we show the mean and Gaussian curvatures per unit area averaged over the last $10^6\unit{fm/c}$ of the run.
We note that for the lowest proton fraction, $Y_p=0.10$ both values are positive, which means several separated convex structures.
For $Y_p=0.20$ the average Gaussian curvature is very close to zero while the average mean curvature is positive.
This is characteristic of convex structures that are on average flat along one direction, such as cylinders.
Meanwhile, both the $Y_p=0.30$ and $Y_p=0.40$ systems have positive average mean curvature and negative Gaussian curvatures
characteristic of network-like structures \cite{PhysRevLett.111.138301,Schuetrumpf:2014aea}. 
We note that for the $Y_p=0.40$ system both curvatures are close to zero, as the system consists mostly of flat plates.
As we shall see in our discussion of observables, Sec. \ref{ssec:Obs}, 
the $Y_p=0.10$ and $Y_p=0.20$ proton fraction simulations exhibit structure factors that resemble those of a liquid phase.
Meanwhile, the $Y_p=0.30$ and $Y_p=0.40$ simulations display Bragg peaks in their structure factor characteristic of a phase with periodic structures.

\begin{table}[h]
\caption{\label{tab:mf1} Average mean ($B/A$) and Gaussian ($\chi/A$) curvatures for the last one tenth of each run.}
\begin{ruledtabular}
\begin{tabular}{D{.}{.}{1.2} D{.}{.}{1.5} D{,}{}{5.5}}
\multicolumn{1}{c}{$Y_p$} & \multicolumn{1}{c}{$B/A (\unit{fm^{-1}})$ }  & \multicolumn{1}{c}{$\chi/A (\unit{fm^{-2}})$ } \\
\hline
0.10 & 0.415(5)    & 1.23(4)  ,\times10^{-2} \\
0.20 & 0.170(1)    & 6.(12.)\,,\times10^{-5} \\
0.30 & 0.071\,8(9) & -1.15(3) ,\times10^{-3} \\
0.40 & 0.012\,7(3) & -3.51(3) ,\times10^{-4} \\
\end{tabular}
\end{ruledtabular}
\end{table}

\subsection{The \textit{waffle} phase}\label{ssec:30}

In this section we focus on systems with proton fractions of $Y_p=0.30$ at a density of $n=0.050\unit{fm}^{-3}$. As seen in the previous section this system has an interesting topology formed of perforated plates we call the ``waffle'' phase. This phase lies in the transition between a phase formed of flat plates, ``lasagna'' phase, and one made up of elongated cylindrical nuclei, ``spaghetti'' phase.

We first discuss simulations performed at a temperature of $kT=1.0\unit{MeV}$ started from a random configuration. To study finite size effects we simulated systems of 51\,200 and 409\,600 nucleons and compared their topologies. We also compare the results obtained from systems that use the artificially decreased screening length $\lambda=10\unit{fm}$ and those obtained from the relativistic Thomas-Fermi approximation, $\lambda_{TF}=13.6\unit{fm}$. All systems were evolved for about $3\times10^7\unit{fm/c}$ in time steps of $2\unit{fm/c}$. 

\begin{figure}[h]
\centering
\begin{overpic}[width=0.5\textwidth]{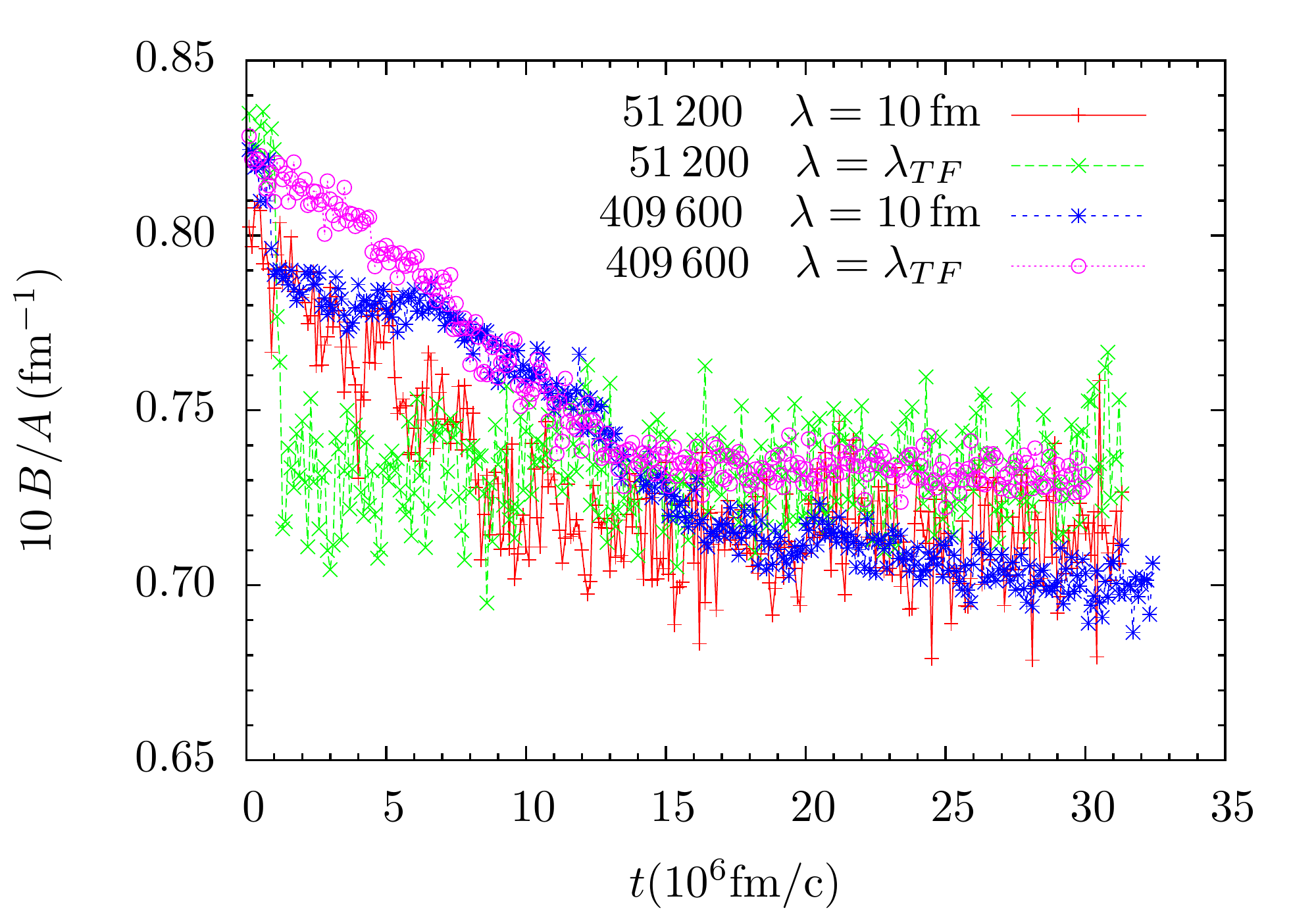}
\put (25,62) {(a)}
\end{overpic}
\begin{overpic}[width=0.5\textwidth]{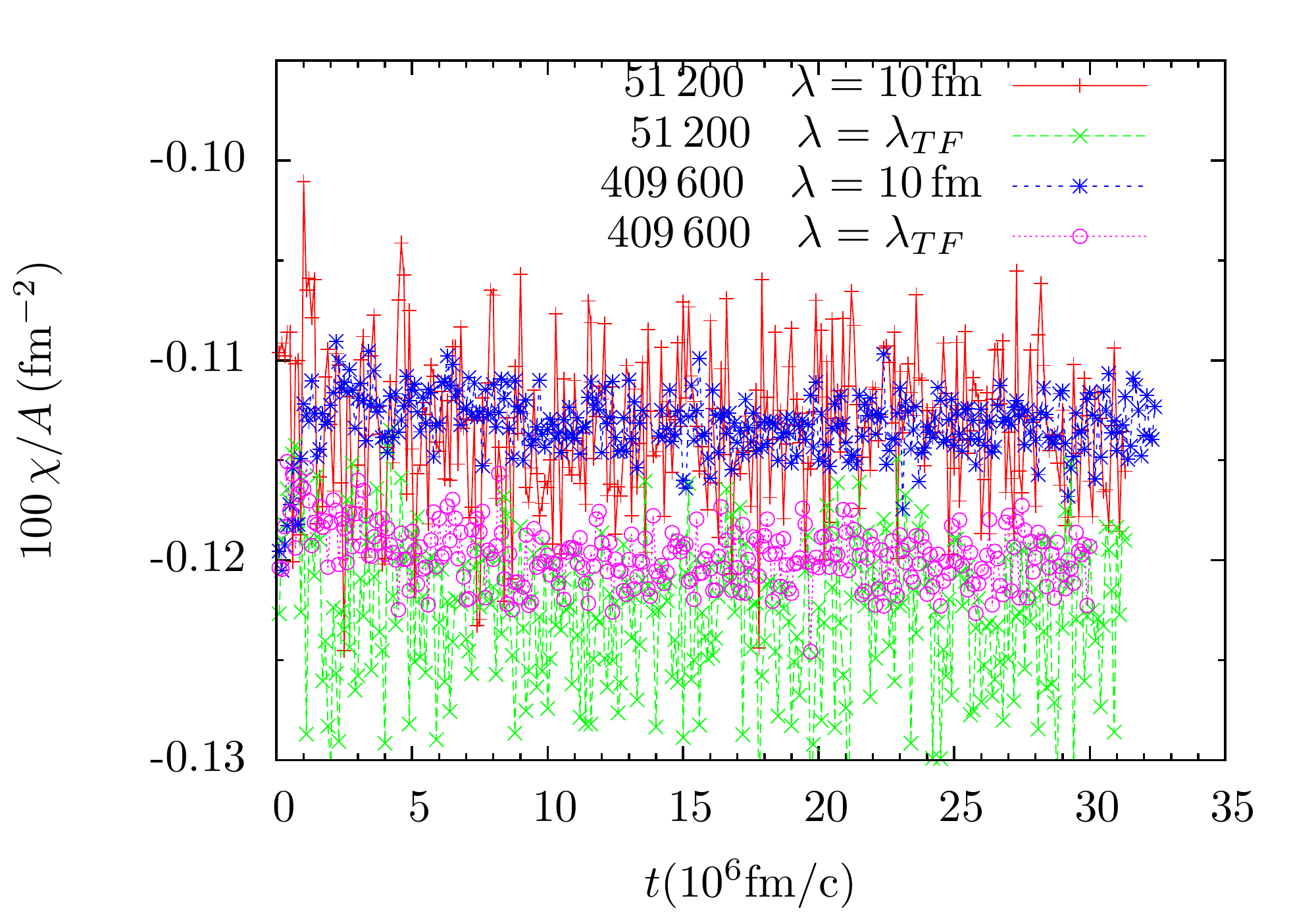}
\put (25,62) {(b)}
\end{overpic}
\caption{\label{fig:bcplot} (Color online) Plots of (a) normalized mean curvature $B/A$ and (b) normalized mean Gaussian curvature $\chi/A$ as a function of simulation time $t$ for four simulations with $Y_p=0.30$, $n=0.050\unit{fm}^{-3}$ and $kT=1.00$ MeV.}
\end{figure}

Comparisons of their topologies can be seen in Figure~\ref{fig:bcplot}. 
The top plot, Figure~\hyperref[fig:bcplot]{\ref{fig:bcplot}(a)}, shows the mean curvature per unit area while the bottom one, Figure~\hyperref[fig:bcplot]{\ref{fig:bcplot}(b)}, shows the Gaussian curvature per unit area of the system as a function of simulation time. 
In Figure~\hyperref[fig:bcplot]{\ref{fig:bcplot}(a)} we see that all systems have initially a mean curvature $B/A\gtrsim0.08\unit{fm}^{-1}$ 
that decreases to $B/A\sim0.07\unit{fm}^{-1}$ as the system evolves.
As expected the 51\,200 nucleon systems equilibrate faster than their 409\,600 counterparts.
Note that here we define equilibrium state as the point where the average mean curvature of the system stops evolving significantly.
In fact, the small systems with screening lengths $\lambda_{TF}=13.6\unit{fm}$ and $\lambda=10\unit{fm}$ seem to
have reached some sort of equilibrium state in about $2\times10^6\unit{fm/c}$ and $10^7\unit{fm/c}$, respectively.
Meanwhile, the larger systems with screening lengths $\lambda_{TF}=13.6\unit{fm}$ and $\lambda=10\unit{fm}$ take somewhat longer to equilibrate.
While the first reaches equilibrium in $2\times10^7\unit{fm/c}$ it is not clear whether the second has reached equilibrium after $3\times10^7\unit{fm/c}$.

While we can infer a time scale for equilibration of the system from the average mean curvature of each simulation the
average Gaussian curvatures only oscillate around an average value soon after the start of the simulation.
As expected the average curvature values depend mostly on the screening length while the deviations from average depend on the number of nucleons in each simulation.
In Table \ref{tab:30} we show the average mean and Gaussian curvatures over the last one tenth of each run.
We see that, even though its not clear whether the larger systems have equilibrated, all values agree well within their standard deviations.

\begin{table}[!htbp]
\caption{\label{tab:30} Average mean ($B/A$) and Gaussian ($\chi/A$) curvatures for the last one-tenth of each run.} 
\begin{ruledtabular}
\begin{tabular}{D{\,}{}{6.0} D{.}{.}{2.1} D{.}{.}{1.7} D{.}{.}{1.7}}
\multicolumn{1}{c}{Size} & \multicolumn{1}{c}{$\lambda (\unit{fm})$} & \multicolumn{1}{c}{$10\,B/A \unit{(fm^{-1})}$ }  & \multicolumn{1}{c}{$100\,\chi/A \unit{(fm^{-2})}$ } \\
\hline
 51\,200 & 10.0 & 0.714(14) & -0.113(3) \\
 51\,200 & 13.6 & 0.735(13) & -0.123(4) \\
409\,600 & 10.0 & 0.700(05) & -0.113(1) \\
409\,600 & 13.6 & 0.731(03) & -0.120(1) \\
\end{tabular}
\end{ruledtabular}
\end{table}

\begin{figure*}[!htbp]
\centering
\includegraphics[width=1.0\textwidth]{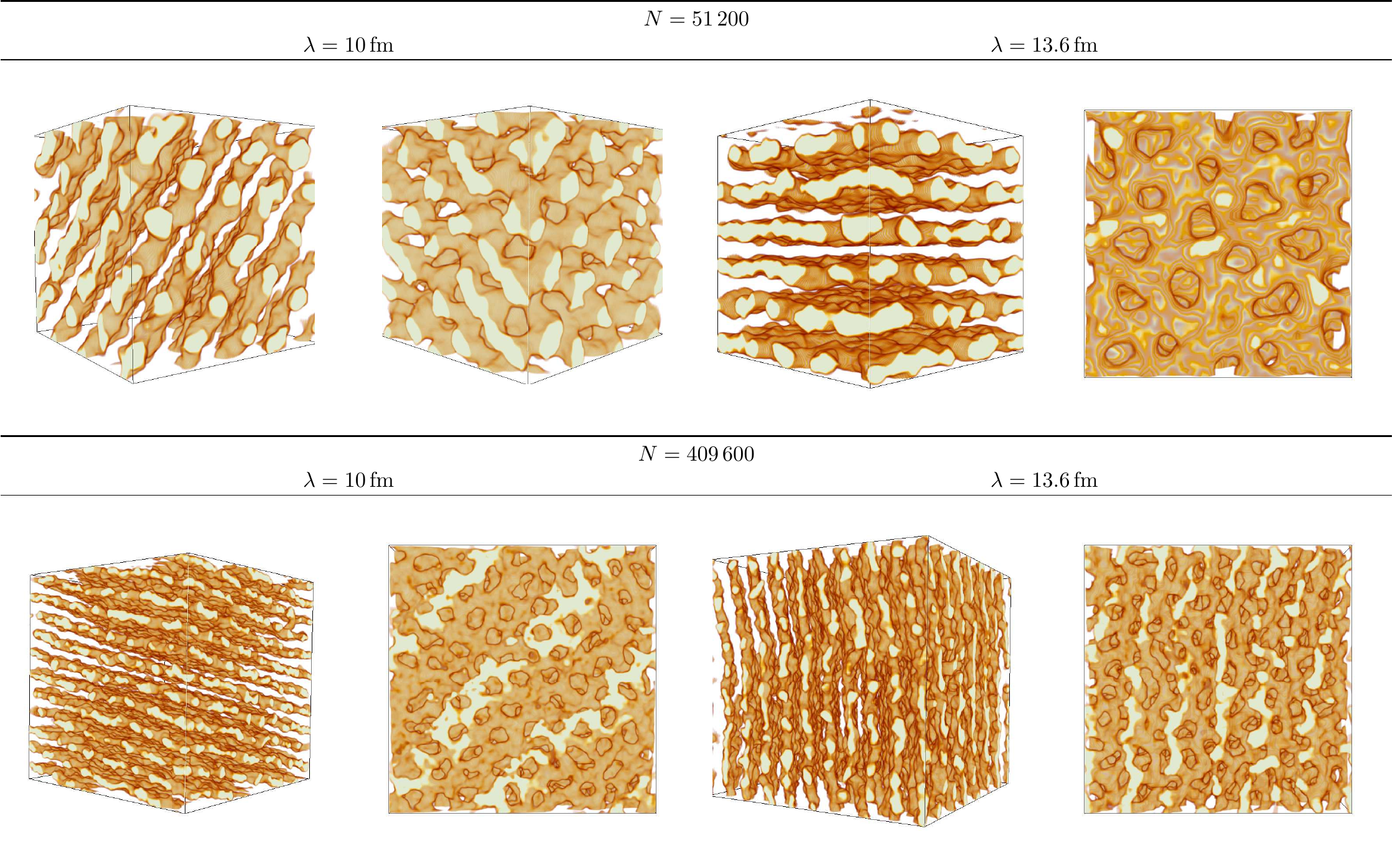}
\caption{\label{fig:30} (Color online) Charge density isosurfaces of runs with 51\,200 (top) and 409\,600 nucleons (bottom).
The two leftmost (rightmost) figures show the final configurations that used a screening length of 10\unit{fm} (13.6\unit{fm}) from different angles.}
\end{figure*}

% In Figure~\ref{fig:30} we show the last configuration of each run from two different angles.
% While three of the runs formed a single perforated plate (this can be checked by analyzing the periodic boundary conditions), 
% the smaller run with $\lambda_{TF}=13.6\unit{fm}$ formed six parallel perforated plates.
In Figure~\ref{fig:30} we show the last configuration of each run from two different points of view. We observe that in every run the final state was formed of perforated plates parallel to each other. Furthermore, in the run with 51\,200 nucleons with $\lambda_{TF}=13.6\unit{fm}$, the plates are also parallel to one of the sides of the box. We also note that even after the long simulation time the larger run with $\lambda_{TF}=13.6\unit{fm}$ exhibited several defects that connected perforated plates aligned along two different directions. This will become clearer in the following discussion of observables, specifically the structure factor $S(q)$.

\begin{figure*}[!htbp]
\centering
\includegraphics[width=1.0\textwidth]{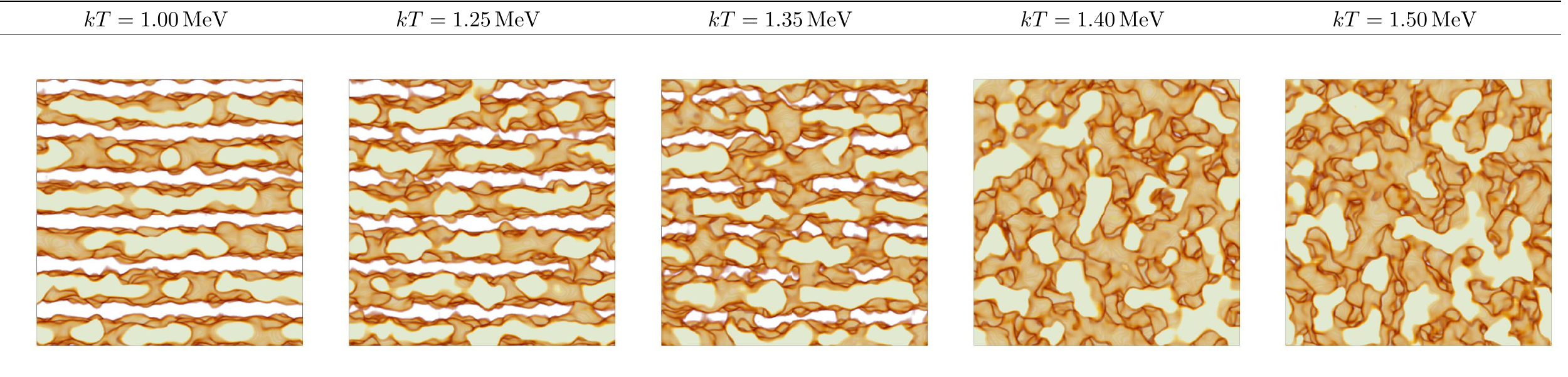}
\caption{\label{fig:hot1} (Color online) Projection along one axis of the charge density isosurfaces of run with 51\,200 and $\lambda=13.6\unit{fm}$ as the temperature is increased from $kT=1.0\unit{MeV}$ to $1.5\unit{MeV}$.}
\end{figure*}

\begin{figure}[!htbp]
\centering
\begin{overpic}[width=0.5\textwidth]{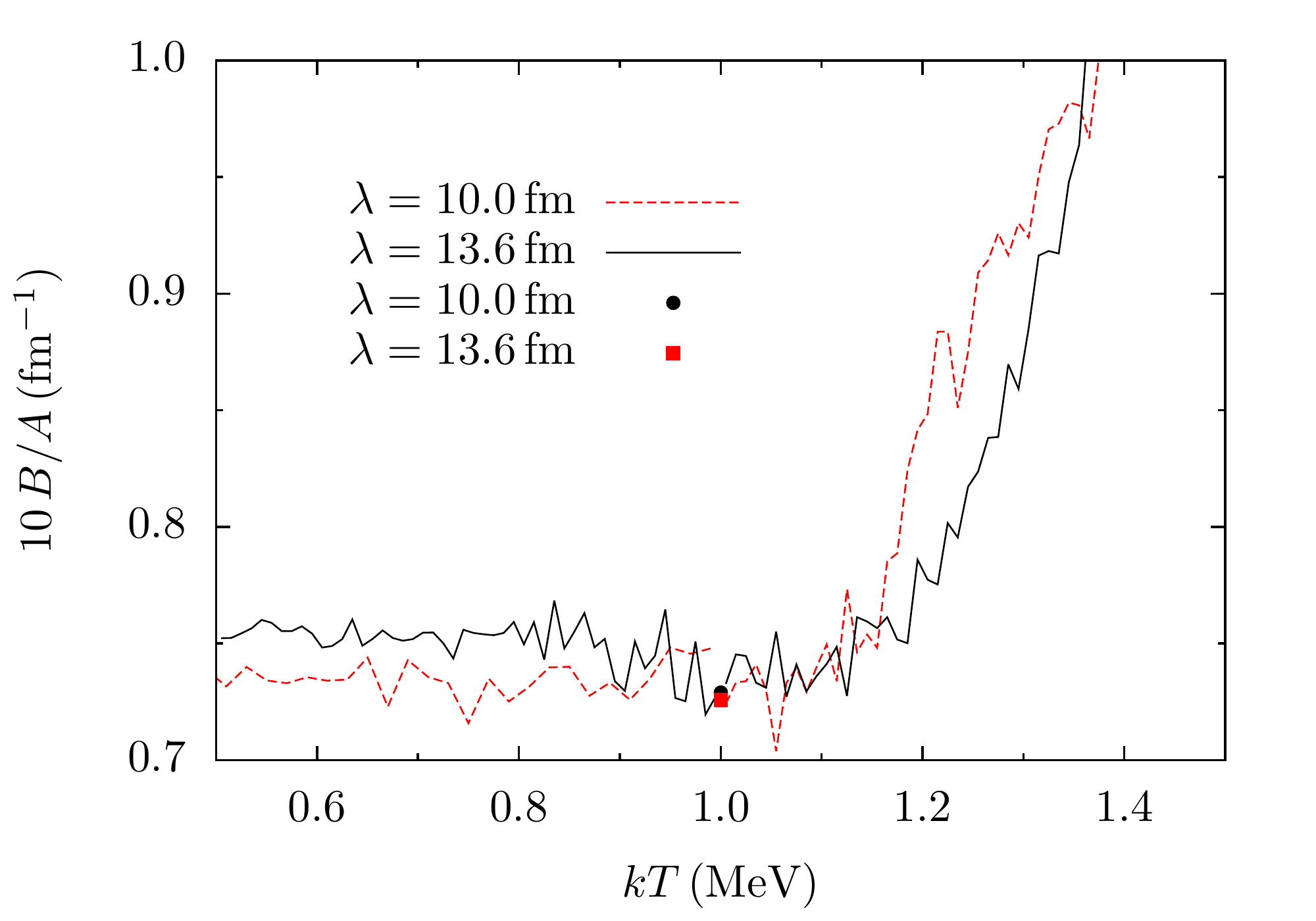}
\put (25,62) {(a)}
\end{overpic}
\begin{overpic}[width=0.5\textwidth]{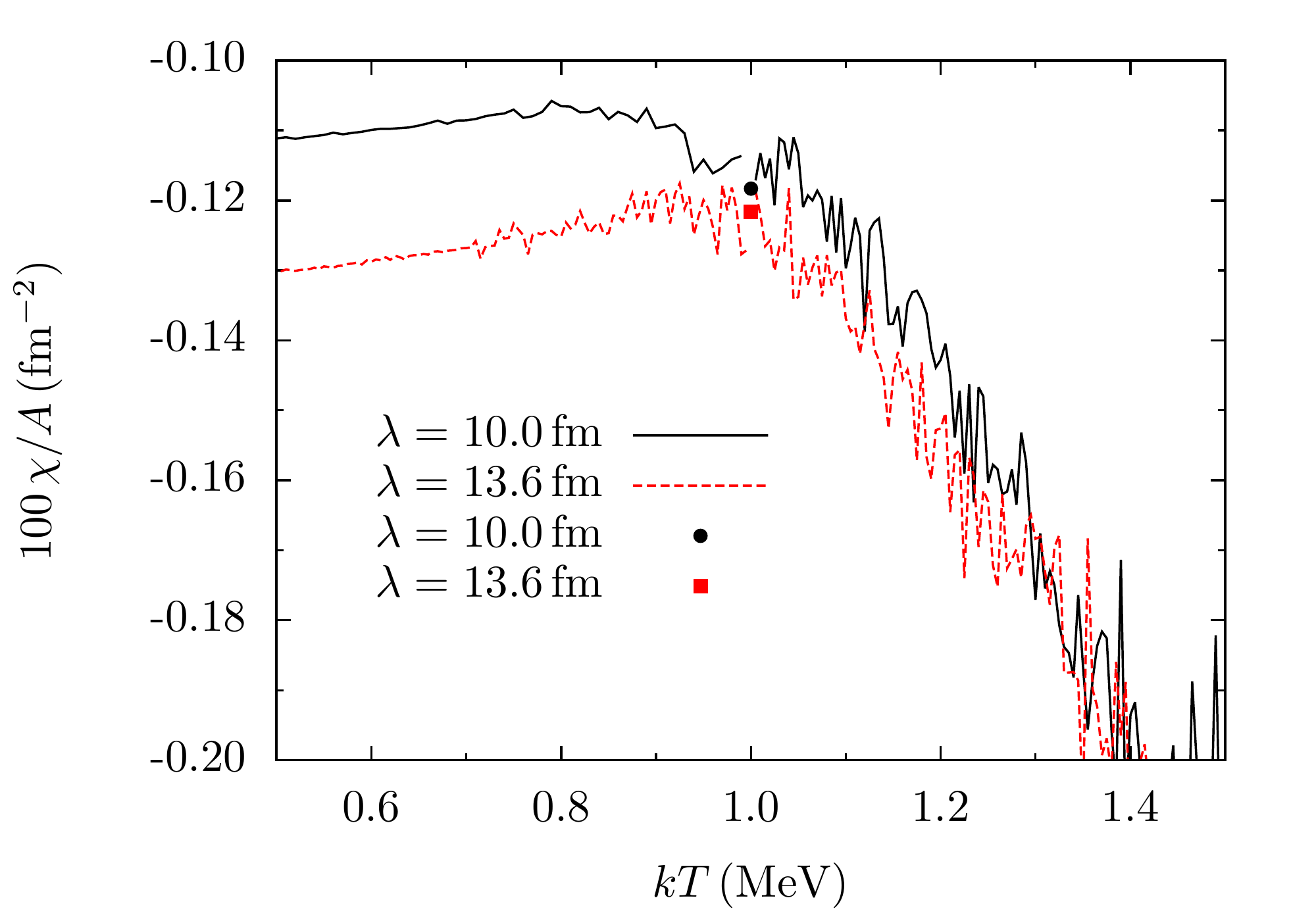}
\put (25,62) {(b)}
\end{overpic}
\caption{\label{fig:plot_kT} (Color online) Plots of (a) normalized mean and (b) Gaussian curvatures as a function of temperature $kT$ for a $Y_p=0.30$ simulation started at a temperature $kT=1.0\unit{MeV}$ and increased or decreased at a rate of $d(kT)/dt=10^{-7}·\unit{MeV/(fm/c)}$. The circles (squares) represent the value for the initial mean curvature for the simulations with $\lambda=10.0\unit{fm}$ ($\lambda=13.6\unit{fm}$).}
\end{figure}

\begin{figure}[h]
\centering
\begin{overpic}[width=0.35\textwidth]{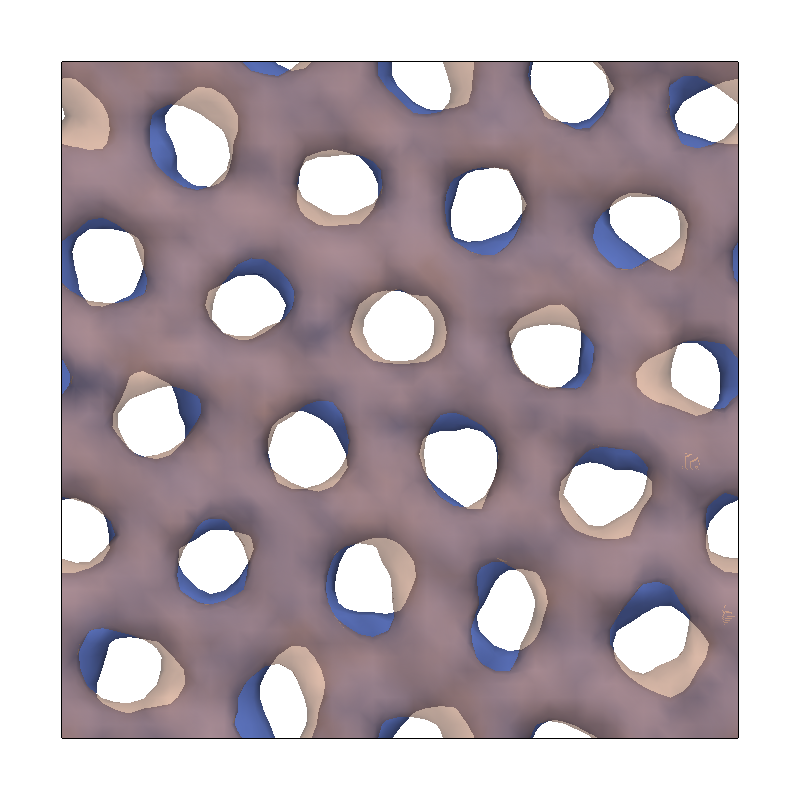}
\put (0,90) {(a)}
\end{overpic}
\begin{overpic}[width=0.35\textwidth]{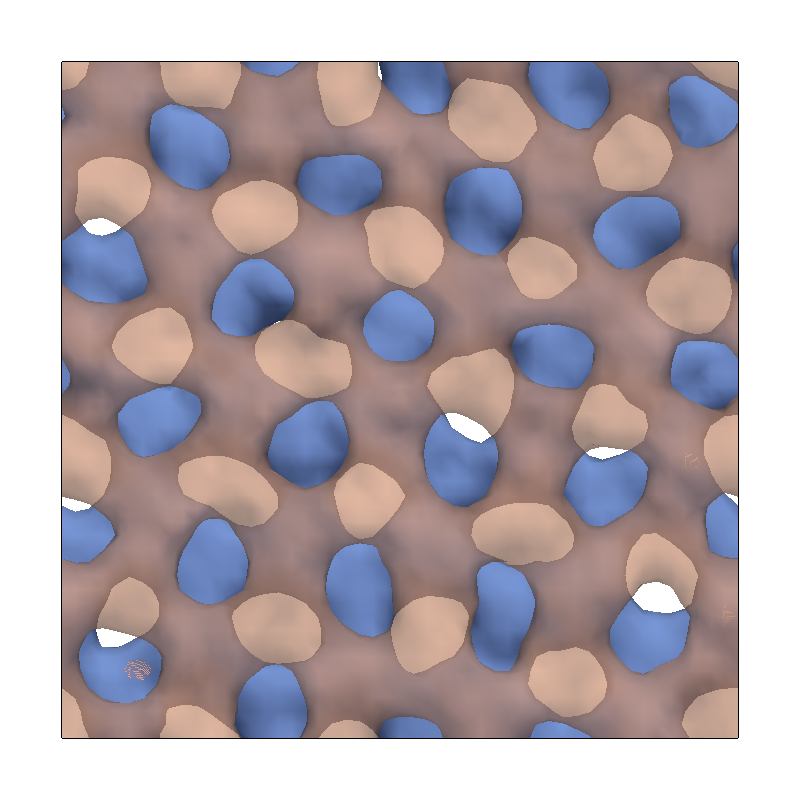}
\put (0,90) {(b)}
\end{overpic}
\caption{\label{fig:cold1} (Color online) Final projection along the $z$ direction of two of the six perforated plates formed in the run with 51\,200 nucleon with proton fraction $Y_p=0.30$ and screening length $\lambda=13.6\unit{fm}$. The simulation was started at $kT=1.0\unit{MeV}$ and cooled to $0.5\unit{MeV}$ (shown). Figure (a) shows two next-nearest neighboring plates, plate 3 (blue) and plate 5 (red) separated by a third plate which is not shown. Figure (b) shows two nearest neighboring plates, plate 4 (blue) and plate 5 (red). The opacity of the plates was decreased so the holes on the blue plates in the back could also be seen.}
\end{figure}

In order to test the stability of these phases we selected the last configuration of the two smaller simulations and slowly increased (decreased) their temperature from $kT=1.0\unit{MeV}$ to $kT=1.5\unit{MeV}$ ($kT=0.5\unit{MeV}$) at a rate of $d(kT)/dt=10^{-7}·\unit{MeV/(fm/c)}$. We then measured the topological characteristics as the system evolved.
We noticed that when the temperature was increased some connections between adjacent plates appeared and at high enough temperatures the pattern of perforated parallel plates merged as the temperature reached $kT=1.3\unit{MeV}$, for an example see Figure~\ref{fig:hot1}. This transition is characterized by a sudden increase (decrease) in the average mean (Gaussian) curvatures away from their values at $kT=1.0\unit{MeV}$, see Figure~\ref{fig:plot_kT}. Meanwhile, when the temperature is decreased the holes in the perforated plates form a structure close to an hexagonal lattice. Note also that this 2D hexagonal lattice of holes is displaced by about half of a lattice spacing in nearest neighbor plates and, thus, is aligned to the holes in next-nearest neighbor plates. Though this happens in both the $\lambda=10\unit{fm}$ and $\lambda=13.6\unit{fm}$ simulations it is easier to see what happens in the latter as the plates are parallel to one of the sides of the box. Therefore, we chose to show only the $\lambda=13.6\unit{fm}$ plates in Figure~\ref{fig:cold1}. It should be clear comparing the two figures that neighboring plates have holes displaced by half of a lattice spacing so that next-nearest neighbor plates have their holes aligned.

\begin{figure*}[ht]
\centering
\includegraphics[width=1.0\textwidth]{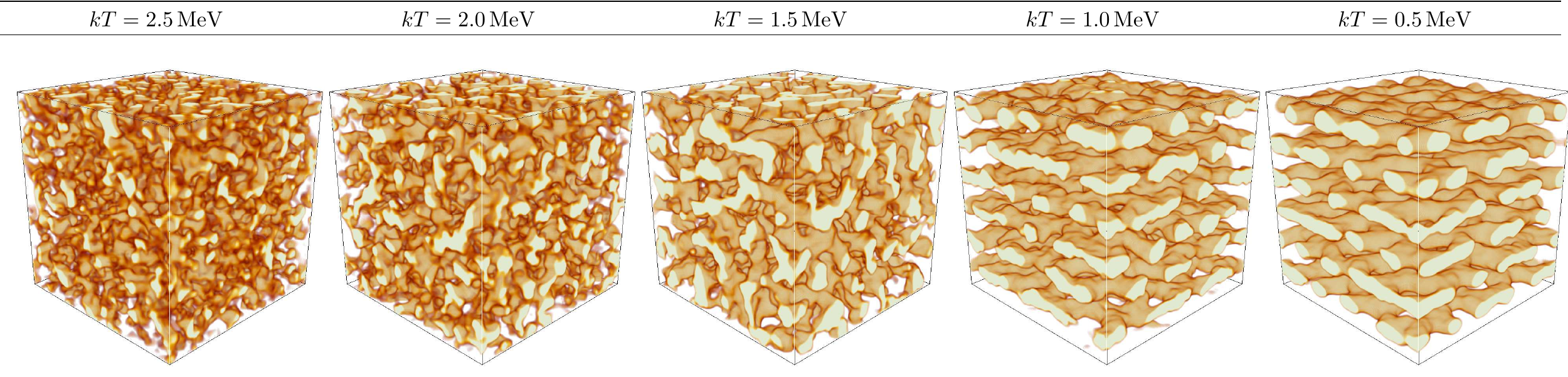}
\caption{\label{fig:kT2} (Color online) Charge density isosurfaces of run with 51\,200 nucleons and $\lambda=13.6$ fm, cooled from $kT=2.5$ to 0.5 MeV.}
\end{figure*}

Besides the simulations described above we also performed two simulations starting at temperature $kT=2.5$ MeV, and cooled at rate $d(kT)/dt = -10^{-7}$ MeV/(fm/c), in order to test the stability of the phase of perforated plates. We used 51\,200 nucleons for both simulations, with $\lambda=10.0$ fm for one and $\lambda=13.6$ fm for the other. We expected that cooling the system slowly enough would allow it to reach an equilibrium state similar to the one found in the constant temperature simulations when it reached $kT=1.0\unit{MeV}$. We expected this since at higher temperatures it is easier for the system to jump the potential barrier that separates states with similar energies. Therefore, once the simulations reached a temperature of slightly below the plate melting temperatures of $kT=1.3\unit{MeV}$ we expected plates to form. However, this only happened for the simulation with $\lambda=13.6\unit{fm}$. In this case the topological characteristics of the system at $kT=1.0\unit{MeV}$ are very similar to those obtained by evolving a random configuration for a long time at $kT=1.0\unit{MeV}$. The systems also look very similar: six parallel perforated plates though their potential energies are slightly different, see Table~\ref{tab:comparison}. This may be due to small differences in the number of nucleons on each plate. The time evolution of this system can be seen in Figure~\ref{fig:kT2}. 

\begin{figure*}[ht]
\centering
\includegraphics[width=1.0\textwidth]{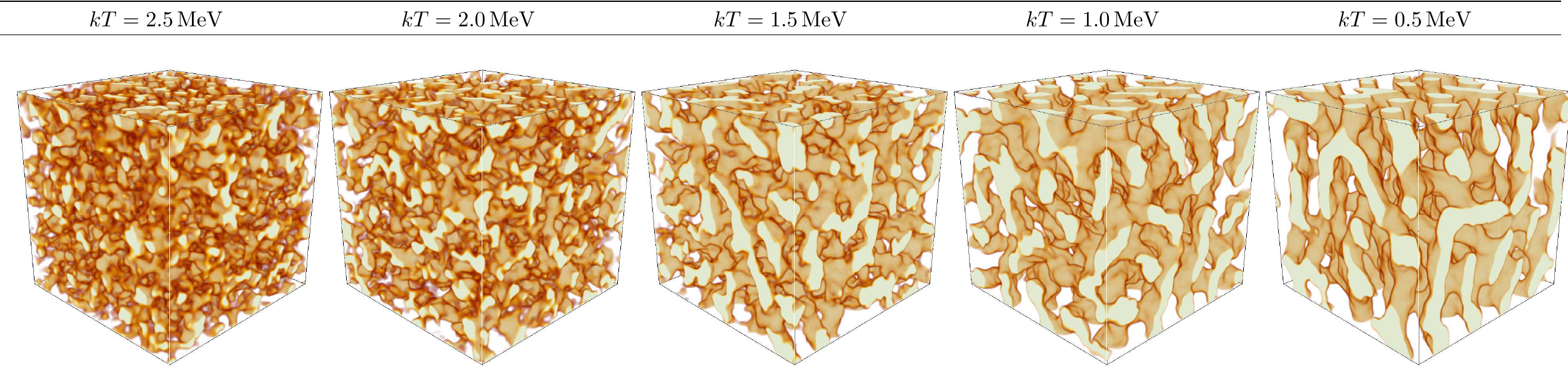}
\caption{\label{fig:kT1} (Color online) Charge density isosurfaces of run with 51\,200 nucleons and $\lambda=10.0$ fm, cooled from $kT=2.5$ to 0.5 MeV.}
\end{figure*}

\begin{figure}[ht]
\centering
\begin{overpic}[width=0.5\textwidth]{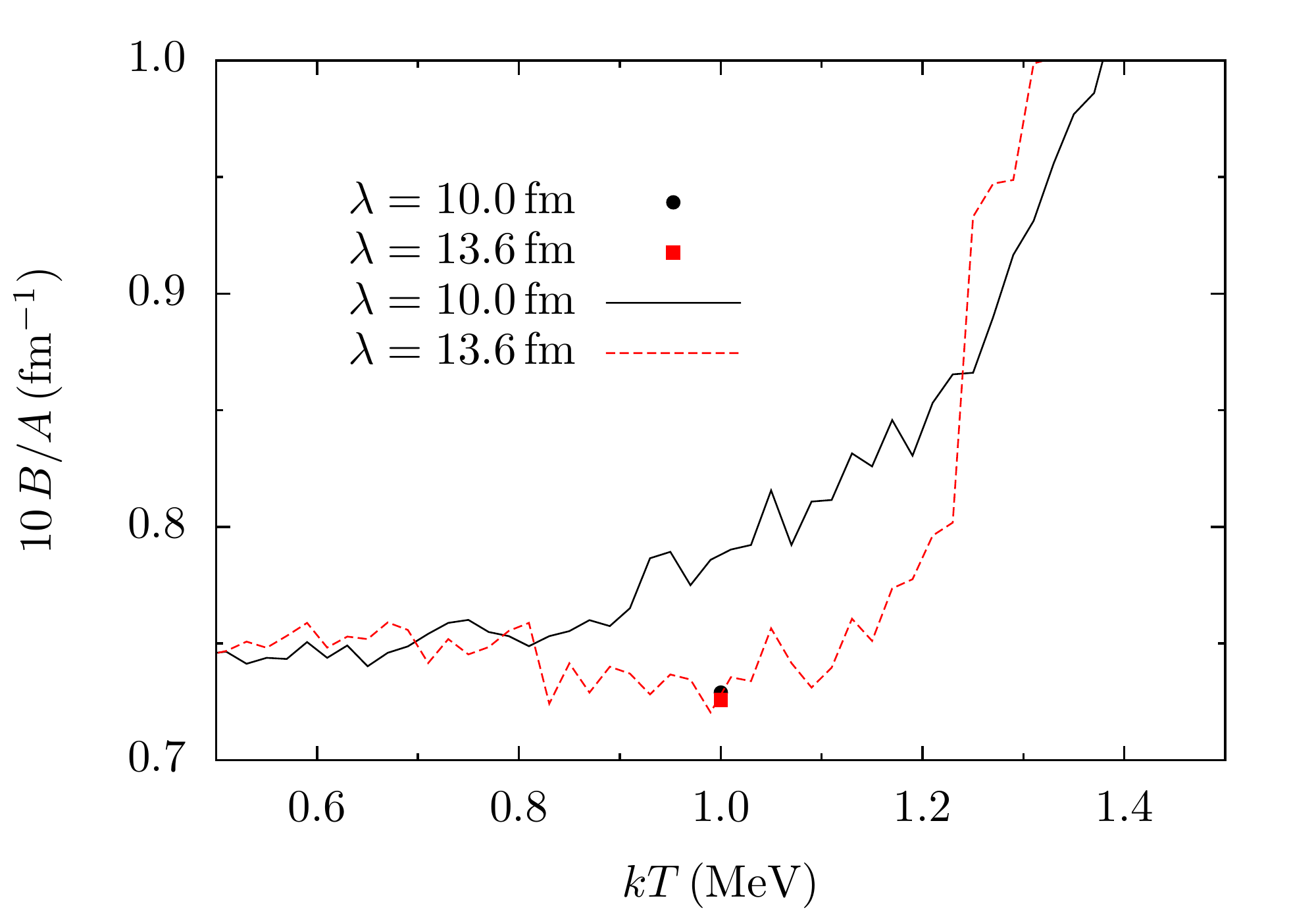}
\put (25,62) {(a)}
\end{overpic}
\begin{overpic}[width=0.5\textwidth]{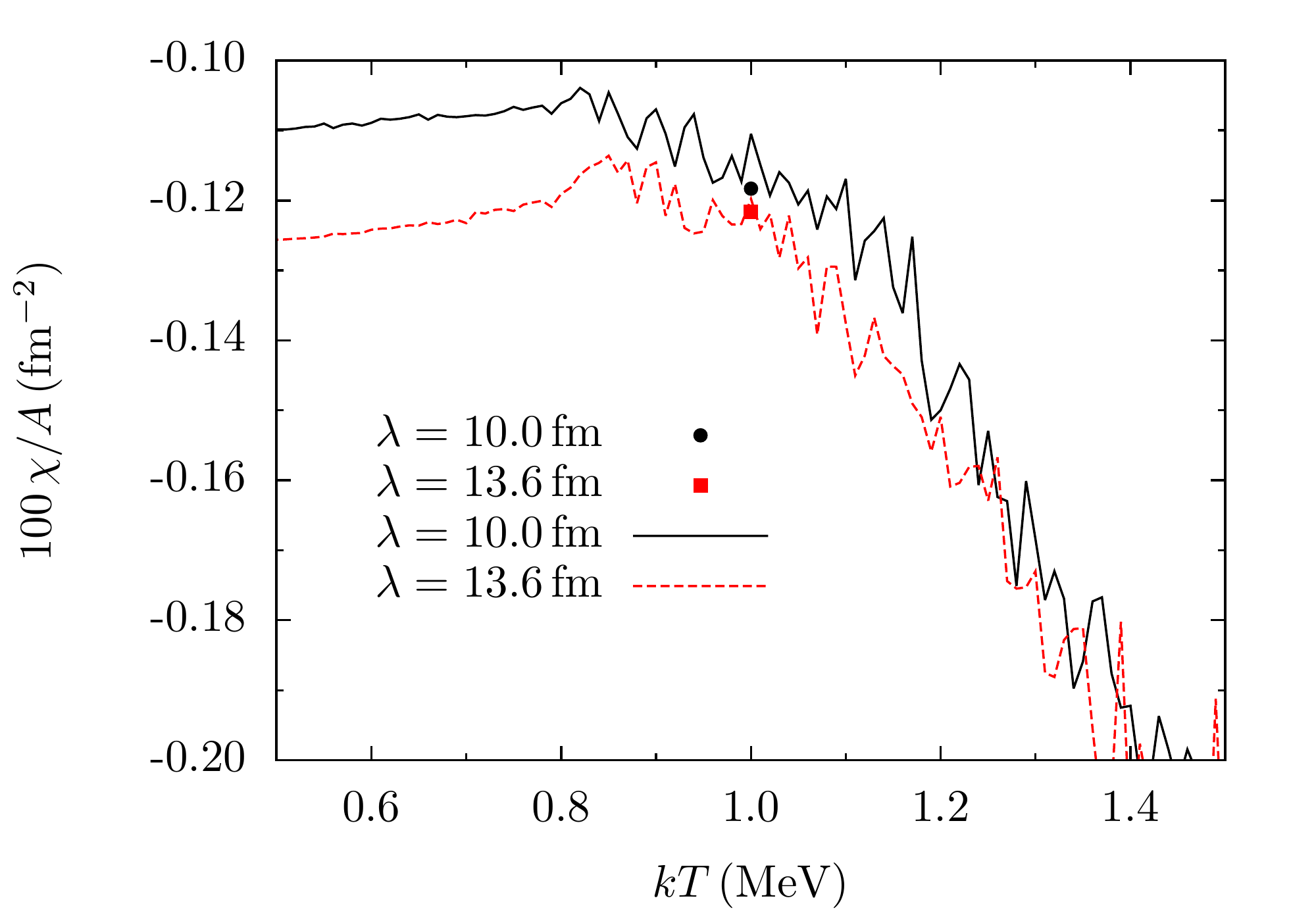}
\put (25,62) {(b)}
\end{overpic}
\caption{\label{fig:plot_kT2} (Color online) Plots of (a) normalized mean and (b) Gaussian curvatures as a function of temperature $kT$ for 51\,200-nucleon systems with $Y_p=0.30$, cooled at rate $d(kT)/dt= -10^{-7}$ Mev/(fm/c) from $kT=2.5$ to 0.5 MeV. One simulation used $\lambda=10$ fm, while the other used $\lambda=13.6$ fm. The circles (squares) represent the initial mean curvatures for the constant $kT=1.0\unit{MeV}$ runs with $\lambda=10.0\unit{fm}$ ($\lambda=13.6\unit{fm}$).}
\end{figure}

Meanwhile, when the run with screening length $\lambda=10\unit{fm}$ reached a temperature of $kT=1.0\unit{MeV}$, down from $kT=2.5\unit{MeV}$, it formed a phase that resembles more several interconnected spaghetti than the perforated plates obtained from evolving an initial random configuration for a long time at a constant $kT=1.0\unit{MeV}$ temperature. The difference in potential energy between the systems at 1.0\unit{MeV} is of the same order of magnitude as the systems run with $\lambda=13.6\unit{fm}$, see Table~\ref{tab:comparison}. This stresses the fact that the difference in energy of systems with significantly different topological characteristics is indeed small. The cooled system may not have reached the waffle phase due to a possible energy barrier once it formed interconnected spaghetti. The evolution of this system can be seen in Figure~\ref{fig:kT1}.
Besides that we also plot the evolution of the topological characteristics of the cooled down systems in Figure~\ref{fig:plot_kT2}. We see that at a temperature of 1.0\unit{MeV} the average curvatures of the system cooled down from 2.5\unit{MeV} are close to the ones obtained from the constant temperature runs for the simulation with screening $\lambda=13.6\unit{fm}$. On the other hand, there are significant differences for the average mean curvature of the two simulations that used a screening length of $\lambda=10\unit{fm}$. These values are also shown in Table~\ref{tab:comparison}.

\begin{table}[!htbp]
\caption{\label{tab:comparison} Topological characteristics and potential energy per nucleon at $kT=1.0\unit{MeV}$ for systems with proton fraction $Y_p=0.30$ obtained from the constant temperature runs and the runs cooled down from $kT=2.5\unit{MeV}$.} 
\begin{ruledtabular}
\begin{tabular}{l D{.}{.}{2.1} D{.}{.}{1.7} D{.}{.}{1.7} D{.}{.}{1.7}}
\multicolumn{1}{c}{Run type} & \multicolumn{1}{c}{$\lambda (\unit{fm})$}  & \multicolumn{1}{c}{$10\,B/A \unit{(fm^{-1})}$}  & \multicolumn{1}{c}{$100\,\chi/A \unit{(fm^{-2})}$} & \multicolumn{1}{c}{$V/N (\unit{MeV})$ } \\
\hline
constant & 10.0 & 0.714(14) & -0.113(3) & -5.6304(1) \\
cooled   & 10.0 & 0.790(10) & -0.116(3) & -5.6246(6) \\
constant & 13.6 & 0.735(13) & -0.123(4) & -2.5807(1) \\
cooled   & 13.6 & 0.736(12) & -0.124(3) & -2.5763(6) \\
\end{tabular}
\end{ruledtabular}
\end{table}

\subsection{Observables}\label{ssec:Obs}

In this section we discuss two observables that can also help us quantify the different pasta structures.
We start with the pair correlation function or radial distribution function (RDF) $g(r)$ and then discuss the structure factor of the pasta shapes $S(q)$.

\begin{figure}[!htbp]
\centering
\begin{overpic}[width=0.35\textwidth]{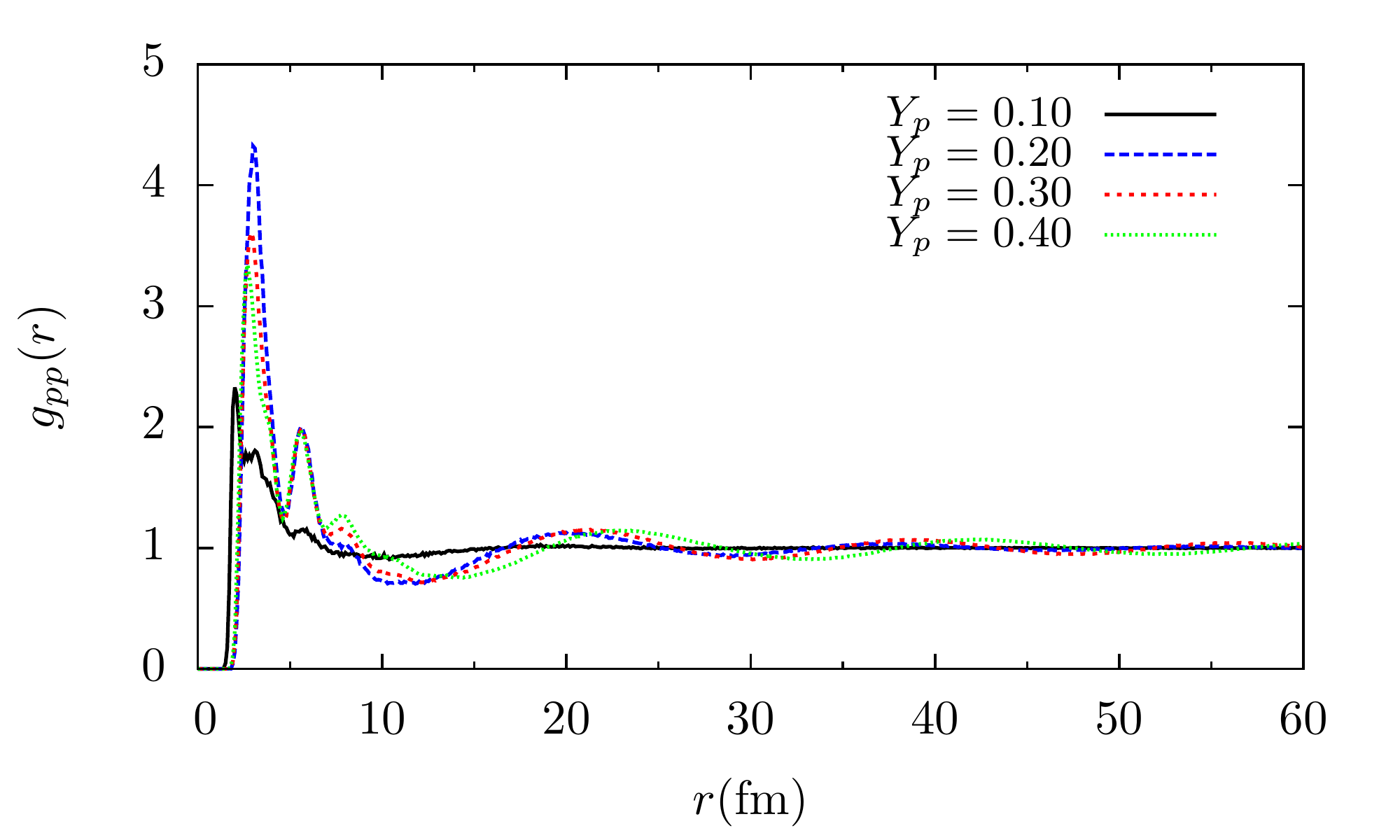}
\put (20,50) {(a)}
\end{overpic}
\begin{overpic}[width=0.35\textwidth]{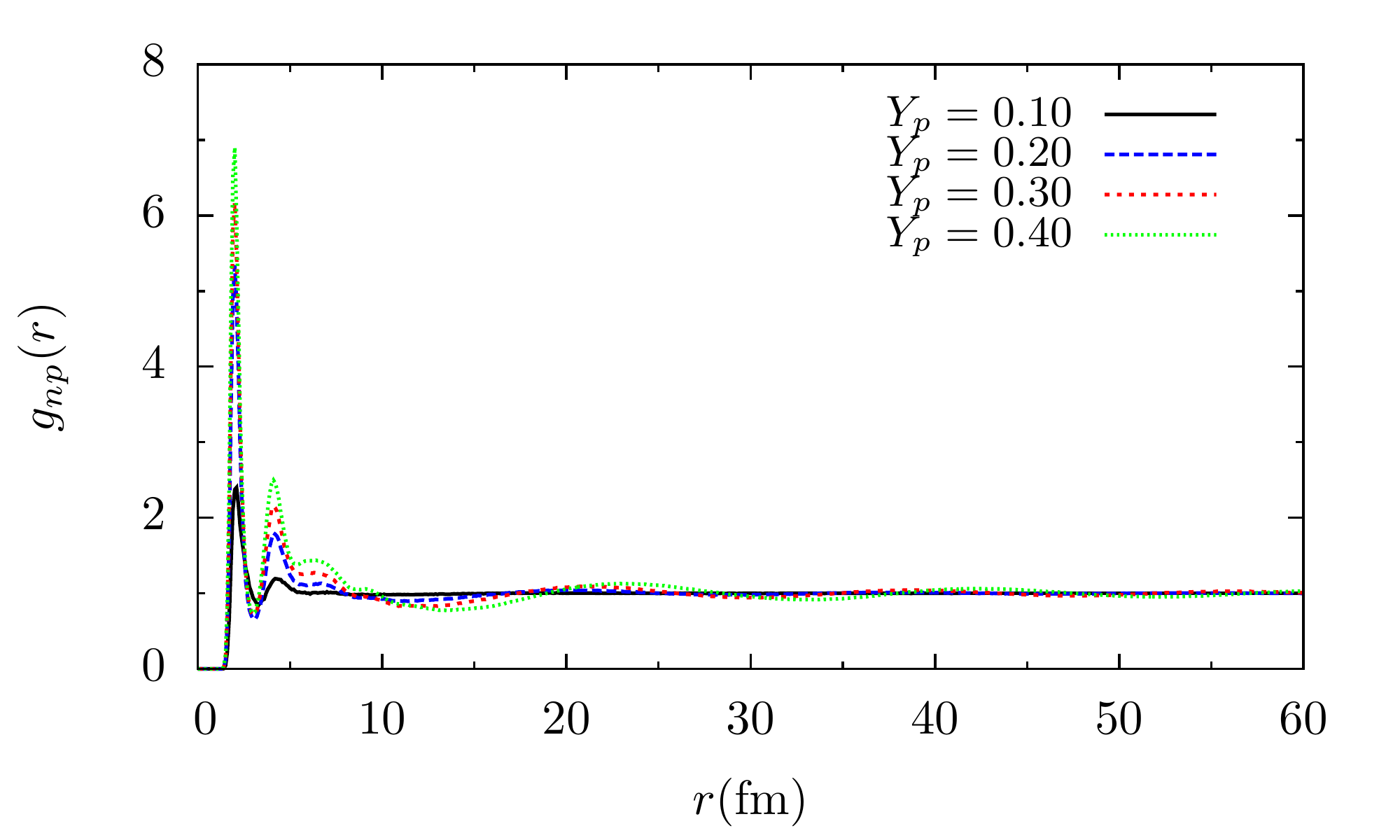}
\put (20,50) {(b)}
\end{overpic}
\begin{overpic}[width=0.35\textwidth]{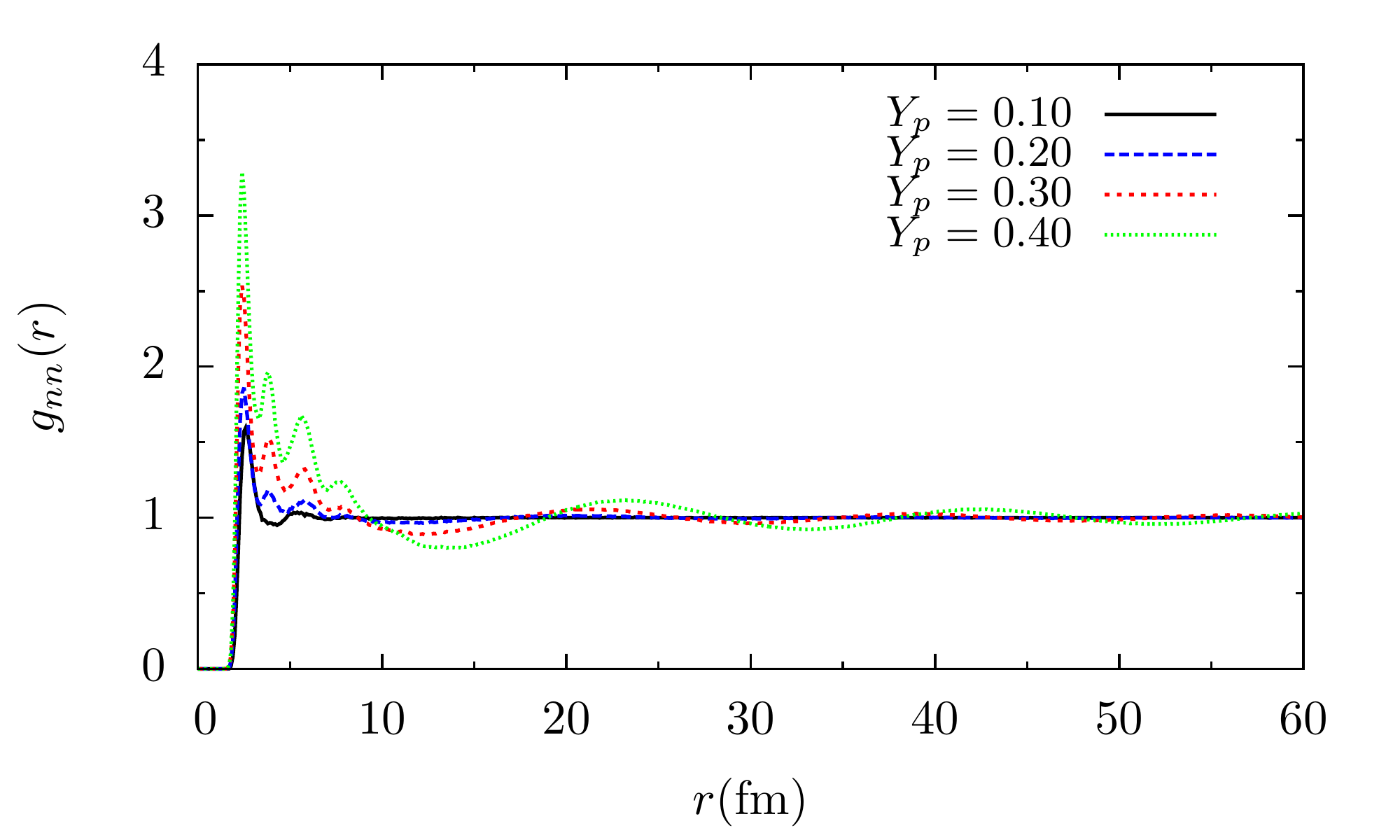}
\put (20,50) {(c)}
\end{overpic}
\caption{\label{fig:gr} (Color online) Radial distribution functions (RDFs) $g(r)$ for (a) proton-proton, (b)
neutron-proton and (c) neutron-neutron pairs for 51\,200 nucleon simulations with a density of $n=0.050\unit{fm}^{-3}$, temperature $kT=1.0\unit{MeV}$, screening length $\lambda=10\unit{fm}$
and proton fractions $Y_p=0.10$, $0.20$, $0.30$ and $0.40$.}
\end{figure}

The RDF $g(r)$ defines the normalized probability of finding a particle of type $a$ at a distance $r$ from a particle of type $b$, \ie,
\begin{equation}
 g_{ab}(r)=\frac{1}{4\pi r^2}\frac{1}{N_aN_b}\sum_{i=1}^{N_a}\sum_{j=1}^{N_b}\langle\delta(\vert\boldsymbol{r}_i-\boldsymbol{r}_j\vert-r)\rangle.
\end{equation}
If $a$ and $b$ are the same type then the sum runs over $i\ne j$ and $N_b=N_a-1$. 
In Figures~\hyperref[fig:gr]{\ref{fig:gr}(a)}, ~\hyperref[fig:gr]{\ref{fig:gr}(b)} and ~\hyperref[fig:gr]{\ref{fig:gr}(c)}, we compare, respectively, $g(r)$ for proton-proton, proton-neutron and neutron-neutron pairs for systems simulated with different proton fractions.
In order to obtain the RDFs we analyzed the positions of all nucleons every 100 time steps over the last $10^6$ time steps of the run.

First we compare the short range behavior of the RDFs. Note that the three systems with higher proton fractions, $Y_p=0.20$, $0.30$ and $0.40$, exhibit similar qualitative behaviors for short range correlations, $r\lesssim10\unit{fm}$; all of their maxima and minima in this region are approximately in the same places regardless of proton fraction, although the heights of these peaks and valleys changes significantly from one system to the next.

The behavior of the RDFs of the low proton fraction system, $Y_p=0.10$, is somewhat different to those of higher proton fractions. For instance, the positions of the first maxima and minima  of the low $Y_p$ system of the proton-proton correlations do not match that of the other systems. Also, the neutron-neutron correlations of this system have only two maxima in the $r<10\unit{fm}$ range while the others have four. We also note that the neutron-neutron and neutron-proton RDF of the $Y_p=0.10$ reach their asymptotic limit of one ($g(r)\rightarrow1$) at $r\sim7\unit{fm}$, while the proton-proton correlations reach this limit at about twice that value. These features may be explained by the fact that the $Y_p=0.10$ system only forms small clusters that are not organized in any particular way and have a large amount of free neutrons in their proximity.

As for the long range behavior, the larger the proton fraction the larger are the oscillations around the asymptotic limit of $g(r)$. This is because the larger proton fraction systems, $Y_p=0.30$ and $0.40$, formed somewhat periodic structures within the simulation volume while the lower proton fraction systems, $Y_p=0.10$ and $0.20$, did not. Also, the long range correlations between proton-proton pairs are stronger than between neutron-neutron and neutron-proton pairs. This is due to two facts. First, there are free neutrons roaming the simulation volume not bound to any cluster and their numbers are larger for the lower the proton fraction. Second, only proton pairs have long range interactions and, therefore, long range correlations that involve a neutron depend on those being bound to nucleon clusters.

Besides the RDFs $g(r)$ we may also obtain the static structure factor $S_a(\boldsymbol{q})$ for nucleons of species $a=n,p$ of the system. 
This quantity is related to the Fourier transform of the pair correlation function $g_{aa}(r)$ \cite{PhysRevC.70.065806}
\begin{equation}
 S_a(\boldsymbol{q})=1+\rho_a\int_V(g_{aa}(r)-1)e^{i\boldsymbol{q}\cdot\boldsymbol{r}}d^3r.
\end{equation}
The structure factor $S_n(\boldsymbol{q})$ of neutrons ($S_p(\boldsymbol{q})$ of protons) can be used to determine the scattering cross section of neutrinos (electrons) by the pasta shapes.
While the neutron structure factor $S_n(\boldsymbol{q})$ may be used to compute neutrino mean-free paths in supernovae and how they are initially trapped, see Reference~\cite{PhysRevC.70.065806}, 
the proton structure factor $S_p(\boldsymbol{q})$ is used to compute thermal conductivity, shear viscosity and electrical conductivity of the pasta, see Reference~\cite{PhysRevC.78.035806}.
To first order, the cross section per neutron of a neutrino of energy $E$ scattered by the pasta is \cite{PhysRevC.70.065806}
\begin{equation}
 \frac{1}{N}\frac{d\sigma}{d\Omega}=S_n(\boldsymbol{q})\frac{G_F^2E^2}{4\pi^2}\frac{1}{4}(1+\cos\theta).
\end{equation}
Here $G_F$ is the Fermi coupling constant, $\theta$ the scattering angle and $\boldsymbol{q}$ the momentum transferred to the system by the incident particle. 
The transferred momentum $\boldsymbol{q}$, the scattering angle $\theta$ and the incident energy are related by
\begin{equation}
 q^2=2E^2(1-\cos\theta).
\end{equation}
Thus, a large structure factor at some transferred momentum $\boldsymbol{q}$ means a large probability that a scattered particle will transfer that momentum to the system.
This occurs whenever the system has a (quasi) periodicity along a direction $\boldsymbol{r}$ such that $\boldsymbol{q}\cdot\boldsymbol{r}\simeq\pm2\pi$.

When calculating the structure factor directly from the Fourier transform of the RDFs obtained from the MD simulations one has to deal with significant finite-size effects as it is difficult to obtain $g(r)$ for $r>L/2$, where $L$ is the size of the simulations cube. This becomes even more troublesome for simulations with higher proton fractions where significant oscillations around the asymptotic limit continue for a distance $r$ much larger than the size of the box. Horowitz \etal\, in Reference~\cite{PhysRevC.69.045804} tried to circumvent that by fitting an exponentially decaying sine function to the tail of $g(r)$. However, this was not helpful in our simulations with $Y_p\geq0.30$. In these cases we noticed that we missed important information about the Bragg peaks in the structure factors that were obtained from the method described next.

As in Horowitz \etal, Reference~\cite{PhysRevC.78.035806}, we calculate the neutron and proton structure factors $S_a(\boldsymbol{q})$ from the density-density correlation function 
\begin{equation}\label{eq:Sq}
 S_a(\boldsymbol{q})=\langle\rho^*_a(\boldsymbol{q})\rho_a(\boldsymbol{q})\rangle
 -\langle\rho^*_a(\boldsymbol{q})\rangle\langle\rho_a(\boldsymbol{q})\rangle.
\end{equation}
The equation above determines the density-density correlations of the neutron and proton densities in momentum space of the system,
\begin{equation}
 \rho_a(\boldsymbol{q})=\frac{1}{\sqrt{N_a}}\sum_{i=1}^{N_a}e^{i\boldsymbol{q}\cdot\boldsymbol{r}_i}.
\end{equation}
In order to avoid finite size effects due to the finite simulation volumes we only take into account transferred momenta $\boldsymbol{q}$ such that
\begin{equation}
 \boldsymbol{q}=2\pi\left(\frac{n_x}{L_x},\frac{n_y}{L_y},\frac{n_z}{L_z}\right)
\end{equation}
where the $n_i\in\mathbb{Z}$ and $L_i$ is the side of the box along the $i$ direction. This choice should be clear since $e^{i\boldsymbol{q}\cdot\boldsymbol{r}}=e^{i\boldsymbol{q}\cdot(\boldsymbol{r+\boldsymbol{L}})}$ for all $\boldsymbol{L}=(m_xL_x,m_yL_y,m_zL_z)$ with $m_i\in\mathbb{Z}$. Note that since our simulation volumes are cubic all $L_i=L$.
In order to obtain the structure factors we saved the configurations of the 51\,200 nucleon runs every 10 time steps over the last $10^6$ time steps of each run.
For the larger 409\,600 nucleon runs we saved $10^4$ configurations over the last $10^6$ time steps of each run.

\begin{figure}[!htbp]
\centering
\begin{overpic}[width=0.5\textwidth]{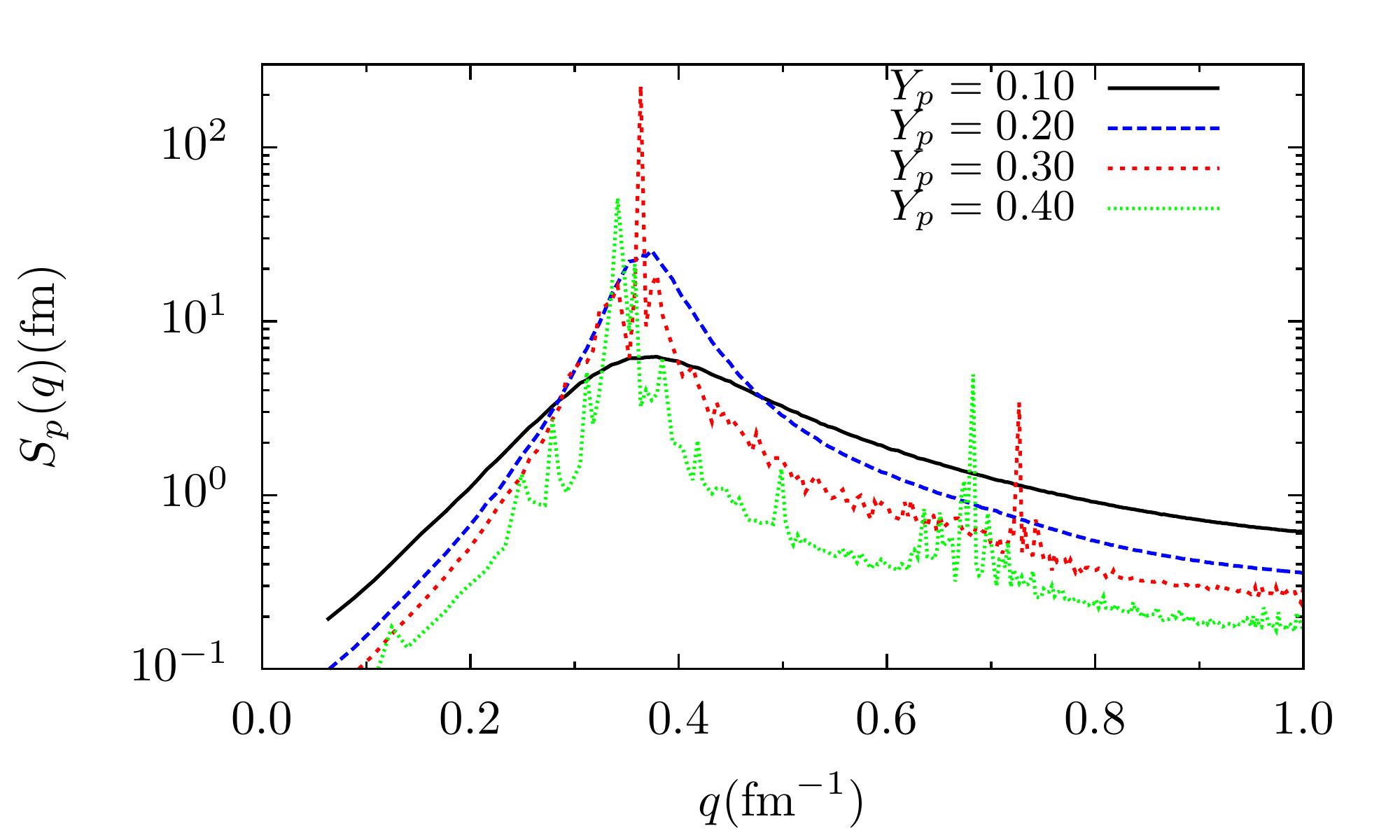}
\put (20,50) {(a)}
\end{overpic}
\begin{overpic}[width=0.5\textwidth]{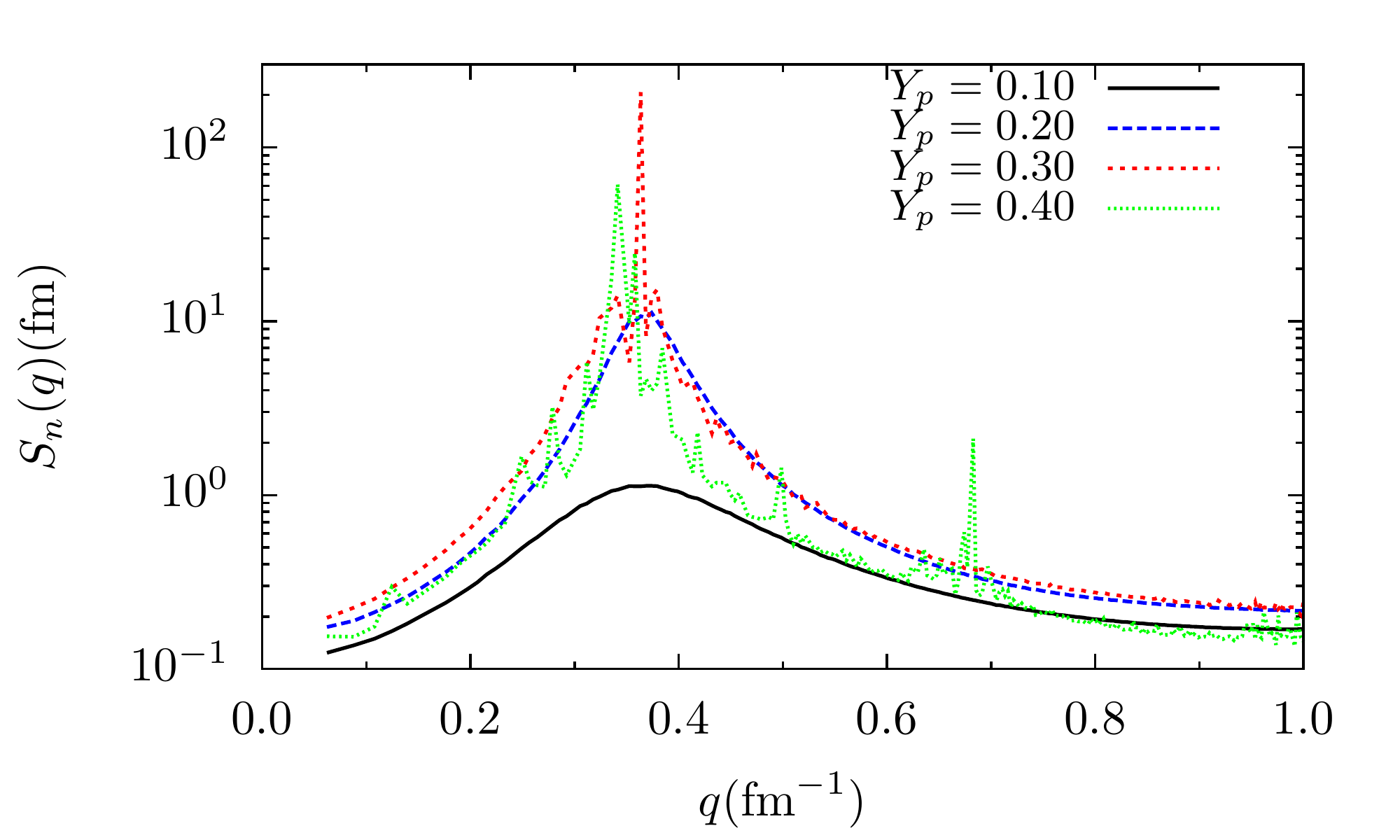}
\put (20,50) {(b)}
\end{overpic}
\caption{\label{fig:sq} (Color online) Angle averaged structure factors $S(q)$ for (a) protons 
and (b) neutrons for simulations with density $n=0.050\unit{fm}^{-3}$, temperature $kT=1.0\unit{MeV}$, screening length $\lambda=10\unit{fm}$
and proton fractions $Y_p=0.10$, $0.20$, $0.30$ and $0.40$.}
\end{figure}

In Figure~\ref{fig:sq} we plot the angle averaged structure factor $S(q)=\langle{S(\boldsymbol{q})}\rangle$ for protons, Figure~\hyperref[fig:sq]{\ref{fig:sq}(a)}, and neutrons, Figure~\hyperref[fig:sq]{\ref{fig:sq}(b)}, for the four simulations discussed in Section~\ref{ssec:Yp}. First we observe that the two simulations with lower proton fractions, $Y_p=0.10$ and $0.20$, have smooth structure factor curves that are characteristic of liquid-like systems. As seen in Figure~\ref{fig:Yp} neither of these two simulations formed periodic structures within the simulation volume. The peaks near $q=0.36\unit{fm}^{-1}$ arise from the average distance between the clusters formed, approximately $L/6$. The height of the peaks is proportional to the contrast in the proton and neutron densities. Therefore, since the $Y_p=0.20$ simulations formed larger clusters than the $Y_p=0.10$ system and the free neutrons gas between its clusters is less dense its peaks are larger.

Meanwhile, the other two simulations, $Y_p=0.30$ and $0.40$, have diffraction peaks characteristic of periodic or solid-like systems. These diffraction peaks come from the values of $\boldsymbol{q}$ perpendicular to the plates formed in the simulation volume. For example, in the $Y_p=0.30$ simulation at $kT=1.0\unit{MeV}$ the transferred momentum that contributes the most to the Bragg peak is the $\boldsymbol{q}=\pm\tfrac{2\pi}{L}(4,3,3)$. This can be checked by looking at the $Y_p=0.30$ configuration in Figure~\ref{fig:Yp}. Note that starting from one of the plates and moving up along the box one reaches another plate every $L/3$. If one moves along  one of the horizontal axis we see plates separated by $L/3$ (left side of the figure) and $L/4$ (right side of the figure). Thus, $\boldsymbol{q}=\pm\tfrac{2\pi}{L}(4,3,3)$ produces the strongest Bragg peak. Its absolute value, $q=0.363\unit{fm}^{-1}$, can be used to estimate the distance $d=2\pi/q\simeq17.3\unit{fm}^{-1}$ between the plates. Also, one expects that for a transferred momentum $\boldsymbol{q}$ that is double of the first peaks, $\boldsymbol{q}=\pm\tfrac{2\pi}{L}(8,6,6)$, there would be another diffraction peak. Though this happens for the proton structure factor $S_p(q)$, it does not for the neutron structure factor $S_n(q)$. This may be due to how the bound neutrons move in the plates or the free neutrons move between them. Another important point is that there does not seem to be any significant diffraction peaks related to the holes in the plates. This is because at $kT=1.0\unit{MeV}$ the shape and position of the holes is constantly changing. This might not be true for that system at lower temperature where the holes in the plates form a two dimensional lattice. Also, it is likely that at slightly higher temperatures than 1.0\unit{MeV} the diffraction peaks disappear altogether as the systems does not have any visible periodic structures within the simulation volume. 

The structure factor of the $Y_p=0.40$ run exhibits several prominent peaks, the largest one being near $q=0.34\unit{fm}^{-1}$. This peak has significant contribution from four different orientations of $\boldsymbol{q}$: $\boldsymbol{q}_1=\pm\tfrac{2\pi}{L}(5,-2,-1)$, $\boldsymbol{q}_2=\pm\tfrac{2\pi}{L}(5,-2,1)$, $\boldsymbol{q}_3=\pm\tfrac{2\pi}{L}(5,2,-1)$ and $\boldsymbol{q}_4=\pm\tfrac{2\pi}{L}(5,2,1)$. The main contribution is from $\boldsymbol{q}_1$ while the other large contributions likely arise from the defects on the pasta structure. In this case, due to the very low number of free neutrons the diffraction peaks appear even in the neutron structure factor at twice and thrice (not shown) the value of $\boldsymbol{q}=0.34\unit{fm}^{-1}$.

\begin{figure*}[!htbp]
\centering
\begin{overpic}[width=0.80\textwidth]{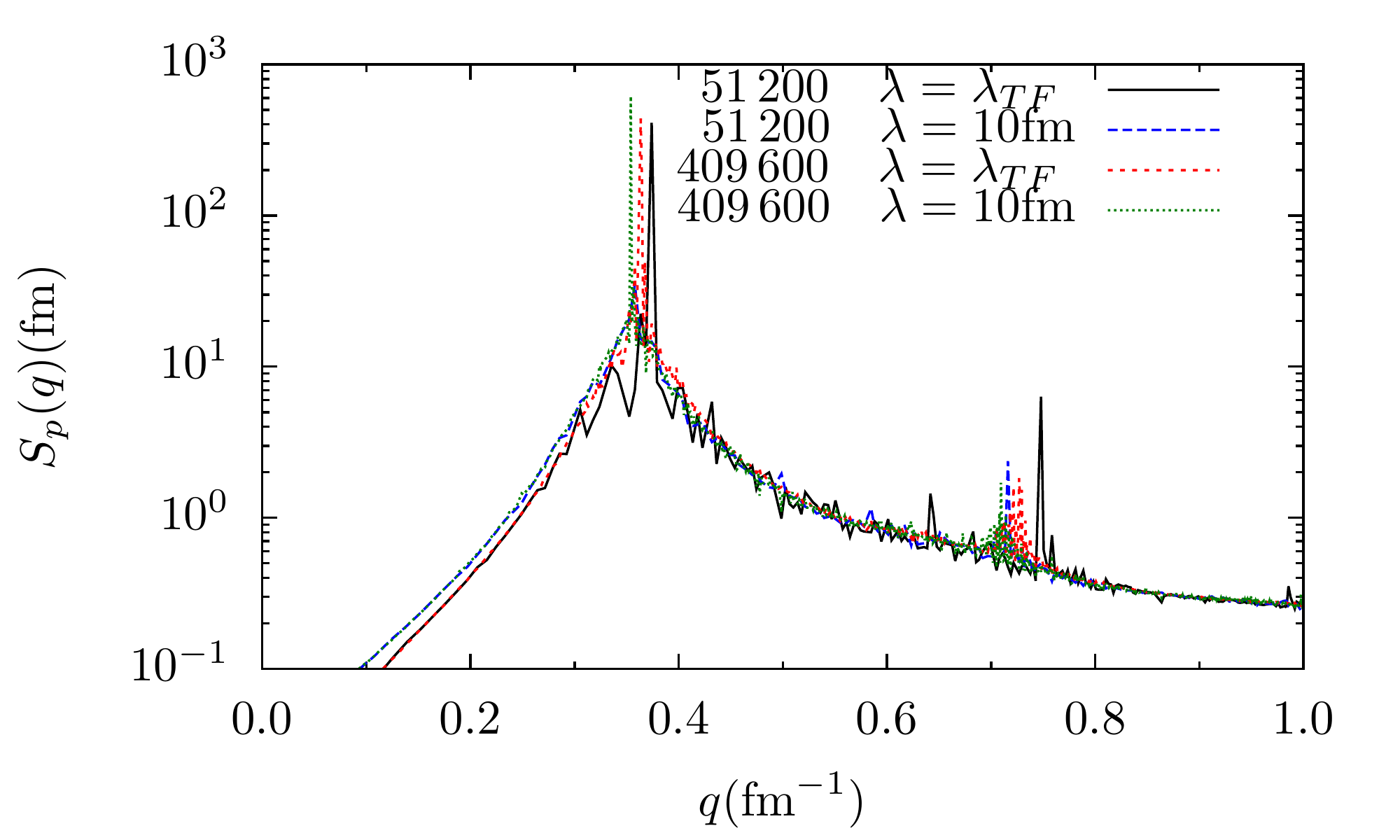}
\put (20,50) {\huge(a)}
\end{overpic}
\begin{overpic}[width=0.80\textwidth]{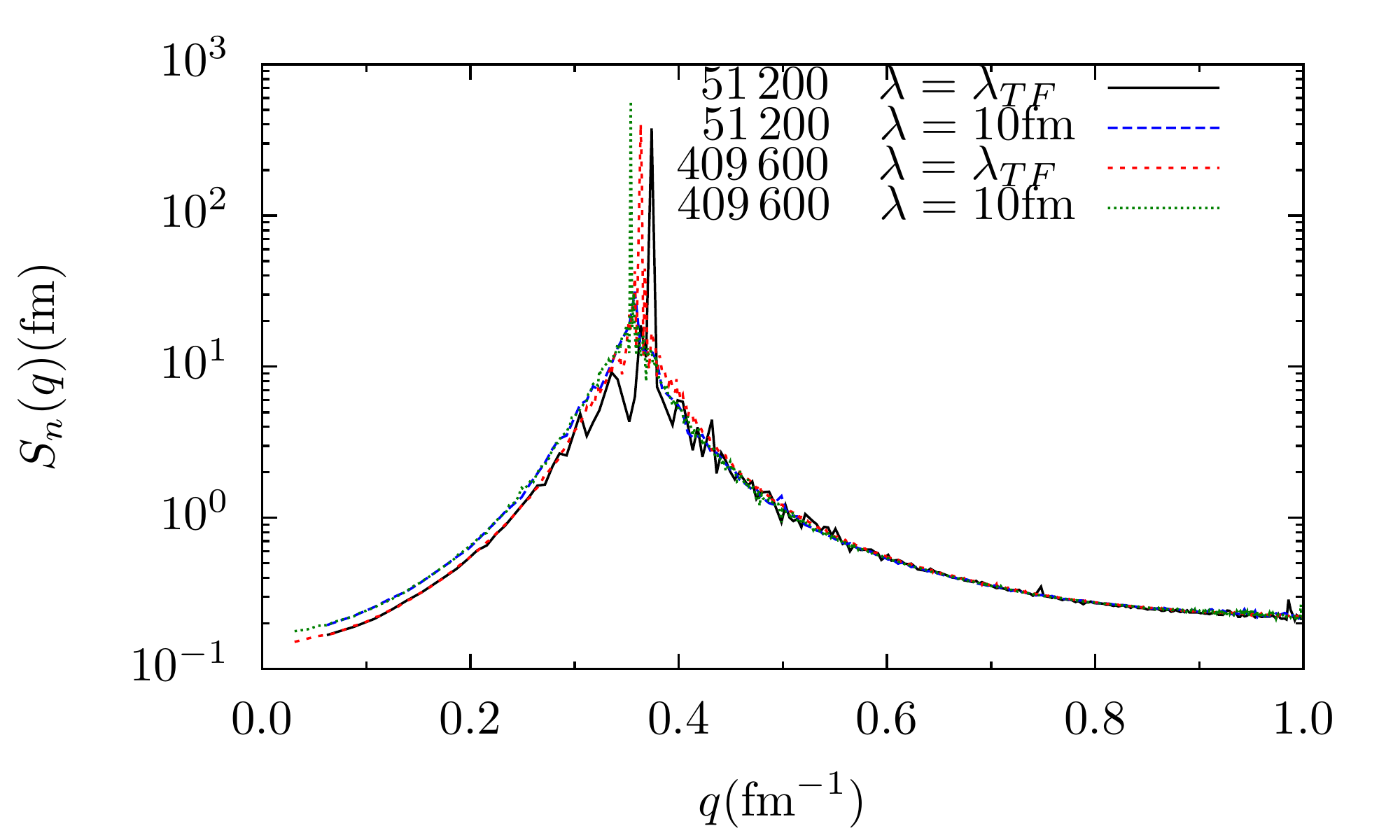}
\put (20,50) {\huge(b)}
\end{overpic}
\caption{\label{fig:sq_30} (Color online) Angle averaged structure factors $S(q)$ for (a) protons and (b) neutrons for simulations with 51\,200 and 409\,600 with density $n=0.050\unit{fm}^{-3}$, temperature $kT=1.0\unit{MeV}$ and proton fraction $Y_p=0.30$.}
\end{figure*}

Another comparison we make is between the structure factors obtained for all of the $Y_p=0.30$ simulations discussed in Section~\ref{ssec:30}. In our comparisons, see Figure~\ref{fig:sq_30}, we first note that finite size effects for the long wavelength limit, $q\lesssim0.30$, of both proton, Figure~\hyperref[fig:sq_30]{\ref{fig:sq_30}(a)},  and neutron, Figure~\hyperref[fig:sq_30]{\ref{fig:sq_30}(b)}, structure factors seem to be well constrained by our simulations. In this region the values for the structure factors only depend on our choice of screening length. On the other hand, for $q\gtrsim0.40$ all the curves are very close to each other provided we ignore the eventual diffraction peaks in the proton structure factors. As discussed above there are no diffraction peaks for the neutron structure factor for $q\gtrsim0.40$ for the $Y_p=0.30$ runs. Though we expect some differences in the structure factors of different runs with different screening lengths we also noted that the number of diffraction peaks and their height and position still depend on the size of the simulation. This implies that, as far as structure factors go, we may need even larger simulations in order to accurately quantify the pattern of diffraction peaks.

\begin{table*}[!htbp]
\caption{\label{tab:peaks} Position $q_{\text{max}}$ and heights, $S_n(q_{\text{max}})$ and $S_p(q_{\text{max}})$, of the highest diffraction peaks for the neutron and proton structure factors and their statistical fluctuations for the $Y_p=0.30$ runs. Also shown are the degeneracy $g_q$ of $q_{\text{max}}$ and the orientations $\boldsymbol{q}_{\text{max}}$ and $\boldsymbol{q}_{\text{min}}$ of the largest and smallest contribution to the diffraction peak at $q_{\text{max}}$.}
\begin{ruledtabular}
\begin{tabular}{D{.}{\,}{3.3} D{.}{.}{2.1} D{.}{.}{1.3} D{.}{}{3.0} D{.}{.}{2.4} D{.}{.}{2.4} c D{.}{.}{3.2} D{.}{.}{3.2} c D{.}{.}{3.2} D{.}{.}{3.2}}
\multicolumn{1}{c}{$N$}&\multicolumn{1}{c}{$\lambda$}&\multicolumn{1}{c}{$q_{\text{max}}$}&\multicolumn{1}{c}{$g_q$}&\multicolumn{1}{c}{$S_n(q_{\text{max}})$}&\multicolumn{1}{c}{$S_p(q_{\text{max}})$}
&\multicolumn{1}{c}{$\boldsymbol{q}_{\text{max}}$}&\multicolumn{1}{c}{$S_n(\boldsymbol{q}_{\text{max}})$}&\multicolumn{1}{c}{$S_p(\boldsymbol{q}_{\text{max}})$}
&\multicolumn{1}{c}{$\boldsymbol{q}_{\text{min}}$}&\multicolumn{1}{c}{$S_n(\boldsymbol{q}_{\text{min}})$}&\multicolumn{1}{c}{$S_p(\boldsymbol{q}_{\text{min}})$}\\
\multicolumn{1}{c}{}&\multicolumn{1}{c}{$\unit{(fm)}$}&\multicolumn{1}{c}{$\unit{(fm^{-1})}$}&\multicolumn{1}{c}{}&\multicolumn{1}{c}{$\unit{(fm)}$}&\multicolumn{1}{c}{$\unit{(fm)}$}&\multicolumn{1}{c}{$\unit{(fm^{-1})}$}&\multicolumn{1}{c}{$\unit{(fm)}$}&\multicolumn{1}{c}{$\unit{(fm)}$}&\multicolumn{1}{c}{$\unit{(fm^{-1})}$}&\multicolumn{1}{c}{$\unit{(fm)}$}&\multicolumn{1}{c}{$\unit{(fm)}$}\\
\hline
 51.200 &10.0 & 0.358 &  48.&  31.3(4.0)  &  34.8 (5.6) & $\pm\tfrac{2\pi}{L}(4,4,1)$ &  440.6 &  476.0 & $\pm\tfrac{2\pi}{L}(4,-4,-1)$ & 3.8 & 4.1 \\
 51.200 &13.6 & 0.374 &  30.& 377.5(34.5) & 411.1(48.1) & $\pm\tfrac{2\pi}{L}(0,6,0)$ & 5566.5 & 6053.0 & $\pm\tfrac{2\pi}{L}(6,0,0)$   & 4.4 & 4.7 \\
409.600 &10.0 & 0.354 & 144.& 570.6(3.8)  & 615.4(3.9)  &$\pm\tfrac{2\pi}{L}(2,2,11)$ &40175.  &43266.  & $\pm\tfrac{2\pi}{L}(11,-2,-2)$& 4.7 & 5.1 \\
409.600 &13.6 & 0.363 &  48.& 403.8(5.6)  & 437.2(14.8) &$\pm\tfrac{2\pi}{L}(0,10,-6)$& 9557.3 &10162.  & $\pm\tfrac{2\pi}{L}(6,-6,-8)$ & 4.9 & 5.4 \\
\end{tabular}
\end{ruledtabular}
\end{table*}

In Table~\ref{tab:peaks} we tabulate properties of the diffraction peaks such as their height, position and which orientations of $\boldsymbol{q}$ contribute the most and the least to the peak. First we observe that in all runs the height of $S_p(q_{\text{max}})$ is about $10\pm2\%$ larger than $S_n(q_{\text{max}})$. Also, the main peak positions are within 1\% of each other for the $\lambda=10\unit{fm}$ runs and 3\% for the $\lambda=13.6\unit{fm}$ runs. However, while the peak heights agree within 10\% for the $\lambda=13.6\unit{fm}$ runs there is a factor of 18 in height difference between the 51\,200 and 409\,600 $\lambda=10\unit{fm}$ runs. The short peak in the small run with $\lambda=10\unit{fm}$ screening length comes from a strong cancellation between the first and second terms in the right hand side of Equation~\eqref{eq:Sq} for the orientation of $\boldsymbol{q}$ that contributes the most to the peak; see the $\boldsymbol{q}_{\text{max}}$ column in Table~\ref{tab:peaks}. In fact, while in this run both terms are within 10\% of each other for both protons and neutrons, in the three other runs the first term is a factor of 10 to 60 larger than the second (not explicitly shown). Finally, we note that the smallest contribution to the peaks are often from orientations $\boldsymbol{q}_{\text{min}}$ such that $\boldsymbol{q}_{\text{min}}\cdot\boldsymbol{q}_{\text{max}}\simeq0$. The only run where this is not the case is the 409\,600 nucleon run with screening length $\lambda=13.6\unit{fm}$. However, even in this case the contribution to $S_n(q)$ and $S_p(q)$ from the orientation orthogonal to $\boldsymbol{q}_{\text{max}}$, $\boldsymbol{q}'=\pm\tfrac{2\pi}{L}(0,10,6)$, is of the same order of magnitude as the contribution from $\boldsymbol{q}_{\text{min}}$: $S_n(\boldsymbol{q}')=5.7\unit{fm}$ and $S_p(\boldsymbol{q}')=6.2\unit{fm}$.

There is a second range in momentum transfer $q$ where diffraction peaks appear for the proton structure factors. These are located at approximately twice in momentum transfer value as $q$ of the largest peak. As before there are also small differences in the number of peaks around the largest peak and in their positions and magnitudes.

\section{Conclusions}\label{sec:Conclusions}

Using the recently upgraded \textsc{IUMD} code and the newly developed \textsc{CubeMD} we studied nuclear systems at a density of $n=0.050\unit{fm}^{-3}$.
First we discussed the differences in topologies (Minkowski functionals) of four 51\,200 nucleon simulations with different proton fractions at a temperature of $kT=1.0\unit{MeV}$. We observed that the system with a proton fraction of $Y_p=0.10$ formed several small deformed nuclei while the $Y_p=0.20$ system formed elongated nuclei that resembled spaghetti. Meanwhile, both the $Y_p=0.30$ and $0.40$ systems formed network-like structures that spread along the whole length of the simulation volume. By calculating the radial distribution function $g(r)$ we observed that the lower the proton fraction of the system the smaller were the long-range correlations. We also noted that proton-proton correlations $g_{pp}(r)$ exhibited oscillations around the asymptotic value of $g_{pp}(r)$ much larger than the neutron-neutron $g_{nn}(r)$ and neutron-proton $g_{np}(r)$ correlations. Also, except for the lowest proton fraction run, $Y_p=0.10$, all runs had a similar qualitative behavior for the short-range correlations. When we examined the structure factor $S(q)$ of the four systems it became evident that the two systems with lower proton fractions exhibited a liquid-like behavior while the two higher proton fraction systems showed diffraction peaks characteristic of periodic structures inside the simulations volume.

For systems of proton fraction $Y_p=0.30$ we first noticed that the time it takes for the system to equilibrate from a random initial configuration at a temperature of $1.0\unit{MeV}$ depended on system size (51\,200 or 409\,600) and screening length ($\lambda=10.0\unit{fm}$ or $\lambda=13.6\unit{fm}$) used. The system that reached equilibrium fastest was the 51\,200 nucleon run with $\lambda=13.6\unit{fm}$ screening length. It did that in about $2\times10^6\unit{fm/c}$. On the other hand, it was not clear whether the 409\,600 system with $\lambda=10.0\unit{fm}$ reached equilibrium after a $3\times10^7\unit{fm/c}$ simulation time. However, it was obvious that all $Y_p=0.30$ systems were converging to the same phase, a stack of perforated parallel plates. Though the plates formed were stable if the system was kept at a constant temperatures $kT\lesssim1.0\unit{MeV}$ they quickly merged at slightly higher temperatures, $kT\gtrsim1.30\unit{MeV}$. Also, while at temperatures of $kT\simeq1.0\unit{MeV}$ the number, position and shape of the holes were constantly changing. Once the system was cooled to slightly lower temperatures, $kT\lesssim0.75\unit{MeV}$, their positions became approximately fixed and their sizes and shapes were uniform, forming a two dimensional hexagonal lattice. Similar phases have been reported elsewhere in the literature for similar densities and proton fractions, see for example the cross-rods in Reference~\cite{PhysRevLett.109.151101} and rod-2 phase in References~\cite{PhysRevC.87.055805,Schutrumpf:2014vqa}. However, those simulations had much smaller simulation volumes and, therefore, the two dimensional lattice structure formed by the perforations in the lattice may suffer from significant finite-size effects. 

Finally we obtained the structure factor for the $Y_p=0.30$ systems of different sizes and screening lengths. We observed that the qualitative behavior of all structure factors were about the same. At low momentum transferred $q\lesssim0.3\unit{fm}^{-1}$ the structure factors for both neutrons, $S_n(q)$, and protons, $S_p(q)$, depended mostly on the screening length used and was almost independent on the system sizes for our runs. This should be clear since in our simulations the long range periodicity comes from the long range repulsive Coulomb forces and, thus, the distance between structures is highly dependent on the strength of the repulsion. At intermediate momentum transfer, $0.30\unit{fm}^{-1}\lesssim{q}\lesssim0.40\unit{fm}^{-1}$ the structure factors had large Bragg peaks caused by coherent scattering from the periodic structures. Their positions and magnitudes, as well as the vector $\boldsymbol{q}$ that contributed most to the peak, was different for one run to the next. While for the $\lambda=13.6\unit{fm}$ runs the magnitude of the peaks were within 10\% of each other and their positions differed by about 3\%, for the  $\lambda=10.0\unit{fm}$ runs the peak positions were within 1\% of each other and their heights differed by a factor of 18. At large momentum transfer, $q\gtrsim0.4\unit{fm}^{-1}$ the curves of $S_n(q)$ and $S_p(q)$ had the same qualitative behavior which was independent of the screening length used in the simulation, though differences would probably appear had the screening lengths been different enough. The proton structure factor showed a second range of peaks at about twice the value of the first diffraction peaks.

Thus, we conclude that we are able to simulate nucleon systems large enough and for enough time for them to appear to equilibrate. We were able to demonstrate with an independent method from others that there is a stable phase of perforated plates for proton fraction of $Y_p=0.30$ at density $n=0.050\unit{fm}^{-3}$ at low temperatures $kT\lesssim1.0\unit{MeV}$. We also showed how to predict qualitatively and quantitatively the diffraction peaks in the structure factor that should affect heat and thermal conductivities in a neutron star crust and the neutrino opacities of supernovae.

\begin{acknowledgments}

We would like to thank Indiana University for time to run our simulations on the \textsc{Big Red II} supercomputer and acknowledge that figures showing isosurfaces were generated using the \mbox{ParaView} software \cite{Paraview}.

This research was supported in part by Lilly Endowment, Inc., through its support for the Indiana University Pervasive Technology Institute, and in part by the Indiana METACyt Initiative. The Indiana METACyt Initiative at IU is also supported in part by Lilly Endowment, Inc.
This research was also supported by DOE grants DE-FG02-87ER40365 (Indiana University) and DE-SC0008808 (NUCLEI SciDAC Collaboration).

% We are grateful to David Reagan at the Advanced Visualization Laboratory - Indiana University for his help with ParaView.  
% We would also like to thank Indiana University for access to the BigRed supercomputer.  
% This research was supported in part by DOE grants DE-FG02-87ER40365 (Indiana University) and DE-SC0008808 (NUCLEI SciDAC Collaboration) 
% and by the National Science Foundation through XSEDE resources provided by the National Institute for Computational Sciences under grant TG-AST100014.

\end{acknowledgments}

%\section{References}
\bibliography{Waffle}

\end{document}